\documentclass{article}

\usepackage{arxiv}

\usepackage[utf8]{inputenc} 
\usepackage[T1]{fontenc}    
\usepackage{hyperref}       
\usepackage{url}            
\usepackage{booktabs}       
\usepackage{amsfonts}       
\usepackage{amsmath}
\usepackage{nicefrac}       
\usepackage{microtype}      
\usepackage{lipsum}
\usepackage{graphicx}
\graphicspath{ {./images/} }
\usepackage{comment}
\usepackage{cleveref}
\usepackage{natbib}
\usepackage{array}
\usepackage{multirow}
\usepackage{tikz}
\usepackage{float}
\usepackage{minted}
\usepackage{caption}
\usepackage{subcaption}
\usepackage{booktabs}
\usepackage{pdfpages}
\usepackage{makecell}
\graphicspath{{Fig/}}

\newcommand{\cmfa}{\texorpdfstring{\textsuperscript{13}C\hbox{-}MFA}{13C-MFA}}
\newcommand{\software}[1]{\texttt{#1}}
\newcommand{\algo}[1]{\textit{#1}}
\newcommand{\cfluxII}{\software{13CFLUX2}}
\newcommand{\cfluxIII}{\software{13CFLUX(v3)}}
\newcommand{\hopsy}{\software{hopsy}}
\newcommand{\pb}{\software{pybind11}}
\newcommand{\Air}{\texttt{Airflow}}

\newcommand{\ECfull}{\textit{E.~coli}}
\newcommand{\EC}[1]{\textsf{EC{#1}}}
\newcommand{\Synfull}{\textit{Synechocystis}}
\newcommand{\Syn}[1]{\textsf{Syn{#1}}}

\crefname{section}{sec.}{sec.}
\Crefname{section}{Sec.}{Sec.}
\crefname{figure}{fig.}{fig.}
\Crefname{figure}{Fig.}{Fig.}

\usepackage{relsize}
\def\ifmonospace{\ifdim\fontdimen3\font=0pt }

\def\C++{%
\texorpdfstring{
\ifmonospace%
    C++%
\else%
    C\kern-.1067em\raise.35ex\hbox{\smaller[2]{++}}%
\fi%
\spacefactor1000}{C++} }

\title{13CFLUX - Third-generation high-performance engine for isotopically (non)stationary \textsuperscript{13}C metabolic flux analysis}

\author{
 Anton Stratmann\textsuperscript{1,2}, Martin Beyß\textsuperscript{1,2}, Johann F. Jadebeck\textsuperscript{1,2}, Wolfgang Wiechert\textsuperscript{1,2}, Katharina N{\"o}h\textsuperscript{1,$\ast$} \\[2ex]
  \textsuperscript{1}Institute of Bio- and Geosciences, IBG-1: Biotechnology,
  Forschungszentrum Jülich, Jülich, Germany \\
  \textsuperscript{2}Computational Systems Biotechnology (AVT.CSB),
  RWTH Aachen University, Aachen, Germany \\[2ex]
  \textsuperscript{$\ast$} {Corresponding author \href{email:k.noeh@fz-juelich.de}{k.noeh@fz-juelich.de}}
}

\begin{document}

\maketitle

\begin{abstract}
\textsuperscript{13}C-based metabolic flux analysis (\textsuperscript{13}C-MFA) is a cornerstone of quantitative systems biology, yet its increasing data complexity and methodological diversity place high demands on simulation software. We introduce \texttt{13CFLUX}(v3), a third-generation simulation platform that combines a high-performance C++ engine with a convenient Python interface. The software delivers substantial performance gains across isotopically stationary and nonstationary analysis workflows, while remaining flexible to accommodate diverse labeling strategies and analytical platforms. Its open-source availability facilitates seamless integration into computational ecosystems and community-driven extension. By supporting multi-experiment integration, multi-tracer studies, and advanced statistical inference such as Bayesian analysis, 13CFLUX provides a robust and extensible framework for modern fluxomics research.\\
\text{Availability:} Sources and containers are provided at
\href{https://jugit.fz-juelich.de/IBG-1/ModSim/Fluxomics/13CFLUX}{https://jugit.fz-juelich.de/IBG-1/ModSim/Fluxomics/13CFLUX} with
Python packages at \href{https://pypi.org/project/x3cflux/}{https://pypi.org/project/x3cflux/}, and Jupyter notebooks to replicate SI results at \href{https://github.com/JuBiotech/Supplement-to-Stratmann-et-al.-Bioinformatics-2025}{https://github.com/JuBiotech/Supplement-to-Stratmann-et-al.-Bioinformatics-2025}.
\end{abstract}


\section{Introduction}
\label{sec:introduction}
Intracellular metabolic reaction rates (fluxes) at steady-state are crucial for quantitatively understanding cellular metabolism. Determining these fluxes in living cells requires a computational approach, wherein unknown fluxes are inferred from data using metabolic models. Of all fluxomics techniques, \textsuperscript{13}C-based metabolic flux analysis (MFA) is considered the most informative. \cmfa{} utilizes data from isotope labeling experiments (ILE) and external rate measurements to estimate fluxes and their uncertainties within the context of metabolic networks~\citep{Niedenfuhr2015}. This technology is well-established in metabolic and bioprocess engineering, and health research, enabling the characterization of microbes~\citep {Long2019}, plants~\citep{Xu2022}, and mammalian cells~\citep {Hogg2023}.

Advances in experimental-analytical techniques have led to various extensions of \cmfa. For instance, integrating data from multiple ILEs\citep{Crown2015}, either from the same or different analytical platforms~\citep{Rahim2022}, or the integration with genome-scale models~\citep{McCloskey2016_GS,Martin2015}, has enhanced the information gain. Developing case-specific labeling strategies~\citep{BorahSlater2023,Mitosch2023} and mass spectrometry methods~\citep{McCloskey2016_MID,Kappelmann2019} have further expanded the scope of \cmfa{}~\citep{Gopalakrishnan2015,McCloskey2016_GS}, while the miniaturization and automation of ILEs on robotic platforms have improved their economic feasibility~\citep{Fina2023}. When combined with rapid quenching protocols, the label incorporation into intracellular metabolites is trackable in small-scale bioreactor systems such as the BioLector~\citep{Nießer2022}, paving the way for a broader applicability of isotopically nonstationary (INST) \cmfa{}~\citep{Noh2006}. These developments have broadened the application of \cmfa, but they have also raised the bar for the robustness and reliable performance of the computational \cmfa{} toolset.

\begin{figure*}[t]
    \centering
    \includegraphics[width=1.0\textwidth]{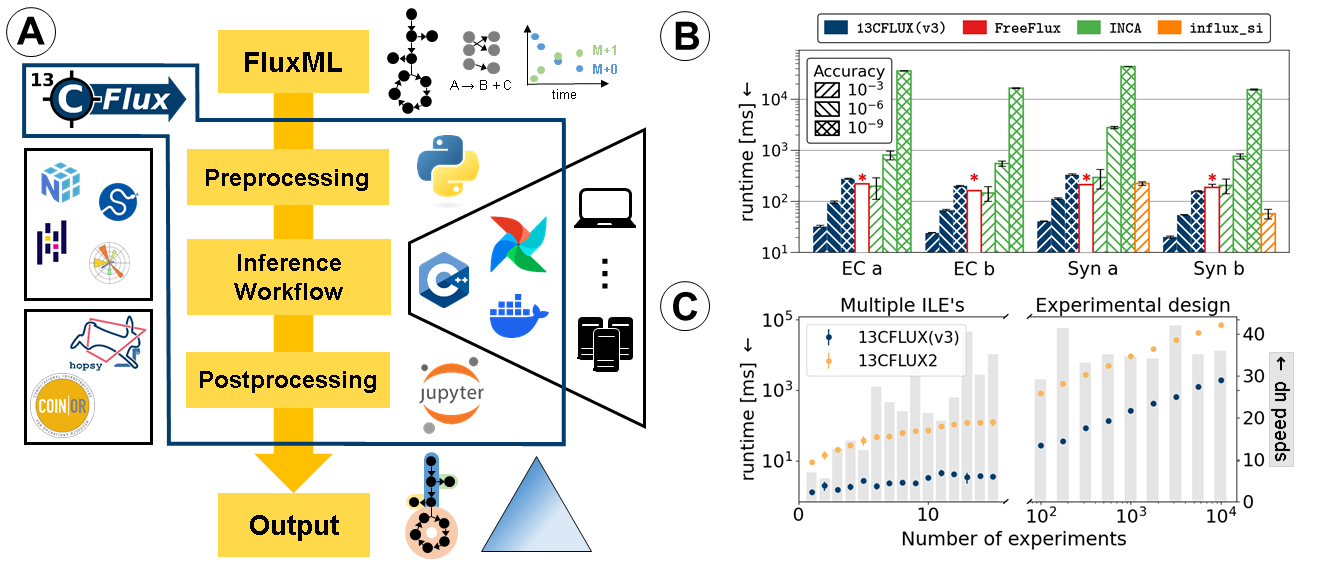}
    \caption{\textbf{(A)} \cfluxIII{} supports flexible, scalable workflows through a high-level Python API that interfaces with the high-performance \C++ simulation backend; Docker containarization ensures portability.
    \textbf{(B)} Performance comparison (single-core) of INST simulation wall clock time for low-, mid-, and high-accuracy calculations (*: numerical accuracy uncontrollable, see SI~\Cref{subsec:benchmark}). Mean and standard deviation over 100 repetitions. 
    \textbf{(C)} Speed-up of \cfluxIII{} vs. \cfluxII{} for simulating multiple isotopically stationary ILEs.}
    \label{fig:overview}
\end{figure*}

The evaluation workflow of \cmfa{} consists of three main steps: experimental design, parameter fitting/optimization, and statistical analysis/uncertainty quantification~\citep{Zamboni2009,Long2019}. To address these steps, numerous methodological developments have emerged, including robust algorithms for ILE design~\citep{Beyß2021}, fast nonlinear optimization~\citep{Lugar2022}, and high-throughput machine learning strategies~\citep{Wu2022}. The classical statistical toolkit has been enriched by Bayesian approaches~\citep{Theorell2017,Backman2023}, allowing researchers to address existing questions and tackle new ones~\citep{Theorell2024}. Consequently, the configuration of \cmfa{} workflows has become increasingly diverse and computationally demanding. Therefore, it is essential that \cmfa{} software supports the flexible composition of analysis workflows to appropriately address the research question at hand while also being streamlined for computational efficiency.

The simulation step is the foundational centerpiece of any evaluation workflow, because it generates isotope labeling data given a metabolic model including atom transitions, parameter values, and a measurement configuration. Reliable and fast simulation of the labeling data enables addressing novel types of questions, such as the impact of model uncertainty on the estimated fluxes~\citep{Theorell2024}. Several \cmfa{} software packages are available~(see SI~\Cref{sec:simulators} for a selection). These use different flux coordinate systems and state-space representations, such as the prominent cumomers or elementary metabolite units (EMU), which determine the properties of the underlying mathematical equations. However, none of these tools cover the full range of now-possible applications, including isotopic stationary and nonstationary \cmfa{} variants with various measurement configurations and multi-isotopic tracers, nor do they select the numerically most beneficial state-space representation. Also, these tools rarely offer the flexibility to accommodate new fitting, statistical, or experimental design approaches.

The open high-performance simulator \cfluxIII{} supports the full range of now possible \cmfa{} scenarios, including INST. Its universality is based on the ability to simulate any desired labeling state of any metabolite within a given model for any input labeling and at any point in time (including $t=\infty$). 
\cfluxIII{} builds on the universal flux modeling language FluxML~\citep{ Beyß2019}, and improves upon the performance of its predecessor \cfluxII~\citep{Weitzel2013}, extending it to INST. Its ground-up new software architecture enables extensible and scalable analyses, facilitating workflow automation, empowering researchers to tackle complex biological questions and applications.
\vfill
\section{Approach and implementation}
\label{sec:approach}
\subsection{Software architecture}
\label{sec:architecture}
The \cfluxIII{} architecture integrates a \C++ simulation backend with a Python frontend for performance and to conveniently leverage third-party Python libraries like \software{NumPy}, \software{SciPy}, or \software{Matplotlib}~(\Cref{fig:overview}A). This cross-language approach, realized using \pb{}~(ver.~3.0.1), compiles the backend and Python bindings into shared libraries accessible to all actively supported Python interpreters (vers.~3.9-13). Advanced exception handling ensures that error and warning messages are passed from the \C++ backend to Python.

Loading a FluxML file in Python creates a simulator object consisting of the dimension-reduced underlying isotope labeling system and data structures tailored to the given \cmfa{} model. This object provides access to simulated labeling data, parameter sensitivities, residuals (variance-weighted difference between simulated and measured data), and gradients, aiding system analysis and flux estimation.

Compared to \cfluxII, the \C++ code has been fully refactored, e.g. by replacing custom matrix/vector operations with those provided by \software{Eigen} (ver.~3.4), thereby reducing the lines of code (LOC) from over 130,000 to less than 15,000 (SI~\Cref{subsec:software_statistics}). This, along with unit testing, enhances maintainability and software quality. The code is written in \C++17~(ISO/IEC 14882) and compiled with standard tools like \software{GCC} or \software{Clang} to highly optimized machine code. \software{CMake} (vers.~3.15-4.1) manages compilation and testing. \cfluxIII{} is deployable as a Python package/wheel e.g. from the Python Package Index (\software{x3cflux}) or as a Docker containers, providing a ready-to-use environment. 

\subsection{\C++ simulation backend}
\label{subsec:simulator_backend}
Battle-proven algorithms are key for achieving simulation performance with high degree of application universality. \cfluxIII{} features two universal state-space representations of isotopic labeling, namely cumomers~\citep{Wiechert2001} and EMUs~\citep{Antoniewicz2007}. 
For a FluxML model, a topological graph analysis and decomposition of the cumomer/EMU isotope labeling balance equations produces dimension-reduced state-spaces (i.e. essential cumomers or EMUs). A heuristic maximizes the reduction by automatically deciding the formulation (SI~\Cref{subsec:labeling_scaling}). The dimension-reduced labeling systems take the form of nonlinearly coupled “cascaded” systems (SI~\Cref{subsec:cascades}), which, depending on the data type, reduce to either algebraic equation systems (AE; isotopically stationary) or ordinary differential equation systems (ODE; INST). Typical system sizes exceed $1000$ dimensions (SI~\Cref{subsec:labeling_dimensions}).

Taking advantage of the systems' sparsity, the AEs are solved using sparse LU factorization via Gaussian elimination with the \algo{SparseLU} algorithm from \software{Eigen}. The \software{SUNDIALS} suite (ver.~6.6)~\citep{Hindmarsh2005,Gardner2022} is used to solve the INST ODEs. In particular, we use a customized version of \algo{CVODE}, a $A(\alpha)$- and $L(\alpha)$-stable multistep Backward Differentiation Formula (BDF) method with step size and order-control, which is suitable for (non)stiff ODE integration ($L$-stability is beneficial as it is robust for integrating stiff ODEs even for large step sizes). The linear system characteristics of the AEs underlying the BDF schemes is leveraged through replacing the iterative \algo{GMRES} algorithm by a 1-step \textit{SparseLU} factorization (SI~\Cref{subsec:taylored-ode-solvers}). Besides \algo{CVODE}, an $L$-stable single-step singly diagonally implicit Runge-Kutta method \citep{Hairer1996} is implemented. All ODE integrators implement adaptive step size control (SI~\Cref{subsec:adaptive-stepsize}), required in all settings where parameter values or design variables vary unpredictably, such as flux estimation or experimental design. Numerical accuracy is validated by comparison with analytical and reference solutions (SI~\Cref{sec:solution-quality}).
The AE/ODE labeling systems and their analytically derived sensitivity systems are solved with the same solvers, enhanced by multi-threaded shared memory parallelization via \software{OpenMP}(vers.~3.0+), which exploits the independence of the sensitivity systems. Labeling states for sets of isotopically labeled substrates, under the same or different measurement configurations, are computed through low-level optimization of the associated AE/ODE systems (SI~\Cref{subsec:multiple_ILEs}).

\subsection{Flexible \cmfa{} workflows}
A key goal in developing \cfluxIII{} was to enable reproducible (automatable, scalable, portable) and transparent workflows without requiring extensive coding expertise. To this end, we adopted an object-oriented approach that abstracts the internal state-space representation and encapsulates the dimension-reduced model in a polymorphic simulator object (\Cref{sec:architecture}). This allows users to specify model parameters (fluxes, pool sizes) and simulation variables (solvers, accuracy, etc.), and to configure tasks like multi-start parameter fitting with minimal code via a high-level Python API. 
For example, multi-start fitting requires a single LOC and switching to a third-party optimizer involves only two changes (SI~\Cref{subsec:interoperability}). This design balances ease of use with flexibility, supporting the integration of external algorithms and enabling the setup of automated production workflows, e.g. via powerful workflow orchestration platforms such as Apache \software{Airflow} (\href{https://airflow.apache.org/}{https://airflow.apache.org/}, SI~\Cref{sec:airflow}).

Documentation and Jupyter notebooks provide templates for both standard and advanced workflows. \cfluxIII{} also issues expressive error messages and warnings at semantic, syntactic, logic, and numeric levels (SI~\Cref{subsec:errors}). Portability is supported through Docker containerization, which decouples workflow development from compute resources. This facilitates reproducible and scalable execution from laptops to high-performance clusters. 

\section{Results}
\label{sec:results}
We demonstrate the utility of \cfluxIII{} via benchmarks against state-of-the-art (SOTA) simulators and, for the first time, Bayesian uncertainty quantification in INST \cmfa{}.

\subsection{Performance benchmark}
\label{subsec:performance}
Benchmarking \cmfa{} simulations includes consideration of the model, measurement configuration (affecting the degree of dimension reduction), state-space representation (cf.~\Cref{subsec:simulator_backend}), and solver accuracy (specific to INST). We compared the simulation wall clock times of \cfluxIII{} with three SOTA simulators (\software{FreeFlux}, \software{INCA}, \software{influx\_si}) for two organisms using published models (\textit{E.~coli} (EC) \citep{Young2014}, \textit{Synechocystis} sp. PCC6803 (Syn) \citep{Wu2023}), two measurement configurations (a,b), and different accuracies (see SI~\Cref{sec:models} for details about the models). Figure~\ref{fig:overview}B and the results in SI~\Cref{sec:simulator_comparison} show that \cfluxIII{} far outperforms the other tools in all categories.
In addition, \cfluxIII{} facilitates scalable simulations of multiple data and parameter sets, achieving $\sim$40 times faster runtimes than \cfluxII{} (\Cref{fig:overview}C). This factor extends to Jacobian computations and significantly speeds up the evaluation of tracer designs (SI~\Cref{subsec:multiple_ILEs_for_ED}).

\subsection{Unlocking Bayesian INST \cmfa{}}
\label{subsec:bayesian-stats}
Bayesian approaches have recently complemented the statistical toolkit for classical \cmfa{}~\citep{Theorell2017,Backman2023,Hogg2023}, but long simulation times have so far hindered their application to Bayesian INST \cmfa{}. Leveraging the performance of \cfluxIII{} together with the efficient MCMC algorithms for linearly constrained problems, we demonstrate Bayesian inference for INST \cmfa{} for the first time. We implemented a Python workflow using \cfluxIII{} as the simulation engine and the specialized library \hopsy{} for MCMC sampling~\citep{Paul2024}. Due to its high-level Python API, integration of the two packages requires only 20 LOCs, with the full analysis workflow implemented under 300 LOCs (SI~\Cref{subsec:hopsy}). The workflow was containerized with Docker and executed on an HPC cluster. Exemplary posterior probability distributions along with further analysis details are provided in SI~\Cref{subsec:BayesApplication}.

\section{Conclusion}
\label{sec:conclusion}
The new simulation engine \cfluxIII{} handles the full spectrum of \cmfa{} scenarios, spanning isotopically stationary and INST analyses, multi-experiment setups, multi-isotope tracers and complex measurement configurations. Its \C++ simulation core ensures efficiency and reliability, while the Python API enables seamless interaction with advanced data analysis tools, workflow customization and automation, thereby supporting both current and future research needs. By uniting performance, reliability, flexibility and openness, \cfluxIII{} establishes a sustainable platform for \cmfa{}, empowering researchers to study complex biological systems and integrate comprehensive datasets to gain quantitative insight into metabolic processes.

\section*{Acknowledgements}
We thank Michael Weitzel for profound insights into \cfluxII{}, Christoph Gerards for contributing to the \cfluxIII{} code, and gratefully acknowledge the computing time on the supercomputer JURECA (grant no. \texttt{loki}).

\section*{Funding}
This work was performed as part of the Helmholtz School for Data Science in Life, Earth and Energy (HDS-LEE); A.S. and J.F.J. received funding from the Helmholtz Association of German Research Centres.

\bibliographystyle{abbrvnat}


\newpage
\appendix
\onecolumn

{
    \centering
    \textbf{Supplementary Information for:}\\[3ex]
    \textbf{\Large 
        13CFLUX - Third-generation high-performance engine \\
        for isotopically (non)stationary \textsuperscript{13}C metabolic flux analysis
        \vspace*{\baselineskip}\\[5ex]
    }
}

\noindent \textbf{Availability and Reproducibility:} \\
Model files and computational scripts are available at \href{https://github.com/JuBiotech/Supplement-to-Stratmann-et-al.-Bioinformatics-2025}{https://github.com/JuBiotech/Supplement-to-Stratmann-et-al.-Bioinformatics-2025}.
A Docker container is provided that allows reproduction \cfluxIII{} results. \\[4ex]

%
%
\section{ Isotope labeling systems}
\subsection{Cascaded isotope labeling and parameter sensitivity equations}
\label{subsec:cascades}

The fractional labeling enrichment in \cmfa{} is described by so-called cascaded labeling differential or algebraic equation systems~\citep{Noh2006}. In the dynamic case, the cascaded systems consist of a sequence of $K$ lower-dimensional linear initial value problems (IVP) for the labeling states ${^k}\mathbf{x}, k=1(1)K$ (cumomer or EMU). Given the parameters, i.e., fluxes $\mathbf{v}$ and pool sizes $\mathbf{X}$ (if applicable), as well as the substrate label composition ${^\text{inp}}\mathbf{x}$, the IVPs are given by
\begin{equation}
    \mathrm{diag}({^k}\mathbf{X}) \cdot {^k}\dot{\mathbf{x}} = {^k}\mathbf{A}(\mathbf{v}) \cdot {^k}\mathbf{x} + {^k}\mathbf{b}(\mathbf{v}, {^0}\mathbf{x}, \dots, {^{k-1}}\mathbf{x}, {^\text{inp}}\mathbf{x}), \quad k=1(1)K
    \label{eqn:cascade-inst}
\end{equation}
with initial labeling states ${^k}\mathbf{x}(t=0) = {^k}\mathbf{x}_0, k=1(1)K$.
For $t \rightarrow \infty$, ${^k}\mathbf{x}$ asymptotically approach constant values and the cascaded IVP simplifies to an algebraic labeling system
\begin{equation}
    \mathbf{0} = {^k}\mathbf{A}(\mathbf{v}) \cdot {^k}\mathbf{x} + {^k}\mathbf{b}(\mathbf{v}, {^0}\mathbf{x}, \dots, {^{k-1}}\mathbf{x}, {^\text{inp}}\mathbf{x}) \quad k=1(1)K
    \label{eqn:cascade-stat}
\end{equation}
The \cmfa{} variants given by Eq.~\eqref{eqn:cascade-inst} and Eq.~\eqref{eqn:cascade-stat} are termed isotopically nonstationary (INST) and isotopically stationary (IST) \cmfa, respectively. 
Notably, the state space representation (cumomer or EMU) affects the dimensions of the cascaded labeling systems: while ${^k}\mathbf{A}$ and $\mathrm{diag}({^k}\mathbf{X})$ are matrices, the inhomogeneous term, ${^k}\mathbf{b}$, and the labeling states, ${^k}\mathbf{x}$, are vector-valued for cumomers, but matrix-valued for EMUs~\citep{WiechertWurzel2001,Young2008}.

When the parameter values in the Eqns.~\eqref{eqn:cascade-inst} and \eqref{eqn:cascade-stat} are unknown, as in the case of parameter estimation, determining the sensitivity of their solution with respect to parameter value perturbations is an important step. Sensitivities in Eq.~\eqref{eqn:cascade-inst} with respect to fluxes and pool sizes are determined by solving the cascaded sensitivity IVPs 
\begin{align}
    \mathrm{diag}({^k}\mathbf{X}) \cdot \dot{(\partial_\mathbf{v}{^k}\mathbf{x})} &= {^k}\mathbf{A} \cdot \partial_\mathbf{v} {^k}\mathbf{x} + \partial_\mathbf{v}{^k}\mathbf{A} \cdot {^k}\mathbf{x} + \partial_\mathbf{v}{^k}\mathbf{b} + \sum_{i=0}^{k-1} \partial_{{^i}\mathbf{x}} {^k}\mathbf{b} \cdot \partial_\mathbf{v} {^{i}}\mathbf{x}, \quad & (\partial_\mathbf{v}{^k}\mathbf{x})(t=0) &= \mathbf{0} \nonumber \\
    \mathrm{diag}({^k}\mathbf{X}) \cdot \dot{(\partial_\mathbf{X} {^k}\mathbf{x})} &= {^k}\mathbf{A} \cdot \partial_\mathbf{X} {^k}\mathbf{x} - \partial_\mathbf{X} \mathrm{diag}({^k}\mathbf{X}) \cdot {^k}\dot{\mathbf{x}} + \sum_{i=0}^{k-1} \partial_{{^i}\mathbf{x}} {^k}\mathbf{b} \cdot \partial_\mathbf{X} {^{i}}\mathbf{x}, \quad & (\partial_\mathbf{X} {^k}\mathbf{x})(t=0) &= \mathbf{0}
    \label{eqn:sensitivities-inst}
\end{align}
for $k=1(1)K$, where $\partial$ is a shortcut for $\mathrm{d} \left/ \mathrm{d}\right.$, and the dependencies of ${^k}\mathbf{A}$ and ${^k}\mathbf{b}$ are omitted for notational brevity. For the IST case, which does not depend on the pool sizes, the sensitivity system reduces to
\begin{equation}
     \mathbf{0} = {^k}\mathbf{A} \cdot \partial_\mathbf{v} {^k}\mathbf{x} + \partial_\mathbf{v}{^k}\mathbf{A} \cdot {^k}\mathbf{x} + \partial_\mathbf{v}{^k}\mathbf{b} + \sum_{i=0}^{k-1} \partial_{{^i}\mathbf{x}} {^k}\mathbf{b} \cdot \partial_\mathbf{v} {^{i}}\mathbf{x}
    \label{eqn:sensitivities-stat}
\end{equation}
\vfill
\subsection{Dimensionality of labeling systems}
\label{subsec:labeling_dimensions}

The dimension of the cumomer and EMU labeling systems increases with the size of the network model~\citep{Weitzel2007}. Analyzing the topological structure of cascaded labeling systems using graph-theoretic concepts significantly reduces the dimension of the systems, yielding essential cumomer and EMU state-space representations \citep{Weitzel2007, Wiechert2013}. Essential cumomer and EMU state-space formulations follow the notion of tracing observed labeling fragments backward to the sources of the label input \citep{Antoniewicz2007r}. Consequently, essential cumomer or EMU state spaces emerge with different dimensionalities and forms (see SI~Table~\ref{tab:dimensionalities}). These, in turn, affect the speed of the computational solutions for labeling and parameter sensitivity systems (see SI~\Cref{subsec:cascades}). For this reason, \cfluxIII{} supports both state-space representations and uses an automated selection heuristic to select the most efficient one.

\begin{table}[!ht]
    \centering
\begin{tabular}{@{}ll>{\raggedright}p{3cm}>{\raggedright}p{3cm}@{\hspace{1cm}}>{\raggedright}p{3cm}>{\raggedright\arraybackslash}p{3cm}@{}}
\toprule
    &   & \multicolumn{2}{l}{\hspace{2.5cm} MS} & \multicolumn{2}{l}{\hspace{1.8cm} MSMS} \\ 
    & Config.  & Cumomer & EMU & Cumomer & EMU \arraybackslash \\ \midrule\addlinespace[10pt]
\multirow{15}{*}{\EC{\_}}  & \textsf{b}  & 151$\times$1, 46$\times$1  & 149$\times$2, 46$\times$3 & 151$\times$1, 46$\times$1 & 151$\times$2, 46$\times$3     \vspace{0.6em} \\
    & \#8  & 152$\times$1, 184$\times$1,120$\times$1, 51$\times$1, 15$\times$1, 2$\times$1 & 149$\times$2, 90$\times$3, 48$\times$4, 11$\times$5, 5$\times$6, 2$\times$7 & 152$\times$1, 184$\times$1, 120$\times$1, 51$\times$1, 15$\times$1, 2$\times$1 & 152$\times$2, 184$\times$3, 120$\times$4, 51$\times$5, 15$\times$6, 2$\times$7 \vspace{0.6em} \\
    & \#20 & 176$\times$1, 268$\times$1, 231$\times$1, 144$\times$1, 74$\times$1, 30$\times$1, 8$\times$1, 1$\times$1 & 150$\times$2, 90$\times$3, 49$\times$4, 20$\times$5, 8$\times$6, 2$\times$7, -- , 1$\times$9 & 179$\times$1, 288$\times$1, 282$\times$1, 221$\times$1, 154$\times$1, 88$\times$1, 36$\times$1, 9$\times$1, 1$\times$1 & 179$\times$2, 288$\times$3, 282$\times$4, 221$\times$5, 154$\times$6, 88$\times$7, 36$\times$8, 9$\times$9, 1$\times$10 \vspace{0.6em} \\
    & \textsf{a}  & 184$\times$1, 289$\times$1, 272$\times$1, 206$\times$1, 145$\times$1, 86$\times$1, 36$\times$1, 9$\times$1, 1$\times$1 & 150$\times$2, 92$\times$3, 49$\times$4, 22$\times$5, 9$\times$6, 2$\times$7, -- , 1$\times$9, 1$\times$10 & 184$\times$1, 289$\times$1, 272$\times$1, 206$\times$1, 145$\times$1, 86$\times$1, 36$\times$1, 9$\times$1, 1$\times$1 & 184$\times$2, 289$\times$3, 272$\times$4, 206$\times$5, 145$\times$6, 86$\times$7, 36$\times$8, 9$\times$9, 1$\times$10 \vspace{0.6em} \\
    & \textsf{full} & 247$\times$1, 536$\times$1, 782$\times$1, 886$\times$1, 837$\times$1, 656$\times$1, 404$\times$1, 183$\times$1, 57$\times$1, 11$\times$1, 1$\times$1  & 247$\times$2, 536$\times$3, 782$\times$4, 886$\times$5, 837$\times$6, 656$\times$7, 404$\times$8, 183$\times$9, 57$\times$10, 11$\times$11, 1$\times$12  & 247$\times$1, 536$\times$1, 782$\times$1, 886$\times$1, 837$\times$1, 656$\times$1, 404$\times$1, 183$\times$1, 57$\times$1, 11$\times$1, 1$\times$1  & 247$\times$2, 536$\times$3, 782$\times$4, 886$\times$5, 837$\times$6, 656$\times$7, 404$\times$8, 183$\times$9, 57$\times$10, 11$\times$11, 1$\times$12 \vspace{0.4em} \\ \hline\vspace{0.1em}\\ 
\multirow{8}{*}{\Syn\_} & \textsf{b}  & 129$\times$1, 138$\times$1, 46$\times$1 & 129$\times$2, 92$\times$3, 46$\times$3 \vspace{0.6em} &  &  \\
    & \textsf{a}  & 129$\times$1, 232$\times$1, 252$\times$1, 176$\times$1, 77$\times$1, 19$\times$1, 2$\times$1 \vspace{0.6em} & 129$\times$2, 92$\times$3, 46$\times$4, 19$\times$5, 14$\times$6, 5$\times$7, 2$\times$8 \vspace{0.6em}  &  &  \\
    & \textsf{full} & 129$\times$1, 232$\times$1, 252$\times$1, 176$\times$1, 77$\times$1, 19$\times$1, 2$\times$1 & 129$\times$2, 232$\times$3, 252$\times$4, 176$\times$5, 77$\times$6, 19$\times$7, 2$\times$8  &  &  \\ \addlinespace[10pt]
    \bottomrule
\end{tabular}
    \vspace{0.2cm}
    \caption{\textbf{State-space dimensionalities of typical \cmfa{} models.} For two exemplary models \EC{} and \Syn{} (see SI~\Cref{subsec:labeling_scaling}), dimensions of essential cumomer and EMU system matrices are listed, resolved by cascade levels. Notice that the dimensionality depends heavily on the measurement configuration (second column) accompanying the \cmfa{} models. \#8 and \#20 give two randomly selected measurement configurations, \textsf{full} the unreduced state-space representation.}
    \label{tab:dimensionalities}
\end{table}
\vfill
\subsection{Efficient labeling state-space formulation for multiple labeling datasets}
\label{subsec:multiple_ILEs}

Using the same \cmfa-model to evaluate multiple isotope labeling datasets that are acquired under very similar conditions has become mainstream~\citep{Long2019}. This requires solving one cascaded system, either in the form of Eq.~\eqref{eqn:cascade-inst} or Eq.~\eqref{eqn:cascade-stat}, per dataset. We showcase a computationally advantageous way to deal with the evaluation of multiple datasets, with the IST \cmfa{} variant at hand (the derivation for the INST case is analogous).

Given $N$ labeling datasets, the cascaded IST equation system are written as
\begin{align}
    \mathbf{0}
    = 
    \begin{pmatrix}
        {^k}\mathbf{A}(\mathbf{v}) & & \\
        & \ddots & & \\
        & & {^k}\mathbf{A}(\mathbf{v})
    \end{pmatrix}  
    \cdot 
    \begin{pmatrix}
        {^k}\mathbf{x}_1 \\
        \vdots \\
        {^k}\mathbf{x}_N
    \end{pmatrix}
    +
    \begin{pmatrix}
        {^k}\mathbf{b}(\mathbf{v}, {^0}\mathbf{x}_1, \dots, {^{k-1}}\mathbf{x}_1, {^\text{inp}}\mathbf{x}_1)\\
        \vdots \\
        {^k}\mathbf{b}(\mathbf{v}, {^0}\mathbf{x}_N, \dots, {^{k-1}}\mathbf{x}_N, {^\text{inp}}\mathbf{x}_N)
    \end{pmatrix}, \, k=1(1)K
    \label{eqn:mile-cascade}
\end{align}
The most costly evaluation in solving Eq.~\eqref{eqn:mile-cascade} is in the inversion of the block-diagonal matrix. Recognizing that the matrices on the diagonal are the same for each dataset, instead of solving the complete system Eq.~\eqref{eqn:mile-cascade}, \cfluxIII{} solves the augmented labeling system
\begin{equation}
    \mathbf{0} = {^k}\mathbf{A}(\mathbf{v}) \cdot {^k}\mathbf{z} + {^k}b(\mathbf{v}, {^0}\mathbf{z}, \dots, {^{k-1}}\mathbf{z}, {^\text{inp}}\mathbf{z}), \quad k=1(1)K
    \label{eqn:mile-cascade-reformulated}
\end{equation}
for the extended labeling states ${^k}\mathbf{z} = \left({^k}\mathbf{x}_1, \dots, {^k}\mathbf{x}_N\right)$.
Thereby, the inversion of ${^k}\mathbf{A}(\mathbf{v})$ has only to be performed once, which reduces computations by $N-1$ fold. Clearly, this reformulation works for cumomers and EMU state-space systems. The resulting speed-up factors are shown in Figure~1C in the main text. Notably, this derivation also applies to cumomer- and EMU-based parameter sensitivity systems in Eqns.~\eqref{eqn:sensitivities-stat} and \eqref{eqn:sensitivities-inst}, where it accelerates experimental design calculations (see SI~\Cref{subsec:multiple_ILEs_for_ED}).

\clearpage
%
%
\section{ Numerical labeling simulation with \cfluxIII{}}
\label{seq:INST_details}
\subsection{Taylored differential equation solvers}
\label{subsec:taylored-ode-solvers}

In \cfluxIII{}, the IVP systems in Eq.~\eqref{eqn:cascade-inst} and Eq.~\eqref{eqn:sensitivities-inst} are solved numerically. For this, a single and multi-step methods are implemented; in particular, a Singly-Diagonally Implicit Runge-Kutta (SDIRK) method of order 4~\citep{Hairer1996}, and Backward Differentiation Formula (BDF) of order 1-5~\citep{Hindmarsh2005} are available. Both SDIRK and BDF solvers require the solution of algebraic equation systems for each cascade level $k=1(1)K$, and for every time step of the integration window. Essentially, per step this amounts to the solution of an algebraic equation of the general form 
\begin{equation}
    \left( \mathbf{I} + \lambda \cdot \mathrm{diag}({^k}\mathbf{X})^{-1} \, \cdot {^k}\mathbf{A}(\mathbf{v}) \right) \cdot \hat{\mathbf{y}} = \mathbf{z}
    \label{eqn:solver-inst}
\end{equation}
where $\lambda$ and $\mathbf{z}$ are solver-specific, and $\mathbf{I}$ is the identity matrix. 
The original implementation of the CVODE solver~\citep{Hindmarsh2005} employs an iterative scheme to solve Eq.~\eqref{eqn:solver-inst}. The BDF implementation in \cfluxIII{}, however, solves the system in one step using lower-upper (LU) factorization, which exploits the system structure of the cascaded labeling systems and the sparsity of the matrices ${^k}\mathbf{A}$, thereby reducing the numerical complexity. The same feature is utilized when solving the sensitivity IVPs in Eq.~\eqref{eqn:sensitivities-inst}. 

\subsection{Adaptive step-size control}
\label{subsec:adaptive-stepsize}

State-of-the-art IVP solvers are equipped with adaptive step-size control, which tries to maintain the accuracy of numerical IVP solutions within a certain tolerance~\citep{Hairer1996}. The adaptive step-size control is based on an estimate of the local numerical (discretization) error, which is determined in each integration step. According to this estimate, the integration step-size is set so that the local error does not exceed a user-specified error tolerance. If this local error exceeds the demanded tolerance level, a smaller step-size is automatically selected. By construction, this also prevents the overall global numerical error of the IVP solution from increasing too much.
Therefore, adaptive step-size control is essential to ensure that the numerical solution process is both reliable and efficient.

SDIRK and BDF schemes implemented in \cfluxIII{} are equipped with adaptive step-size control, with the option of adjusting the relative and absolute integration tolerances, $tol_{rel}$ and $tol_{abs}$, respectively. 
The relative tolerance relates to the error relative to the IVP solution, whereas the absolute tolerance gives the absolute limit, which protects the solution against round-off errors. The cumulative contributions of $tol_{rel}$ and $tol_{abs}$ define the numerical tolerance that is used for estimating the integration step-size for the next time-step. For SDIRK, the step-size control relies on error estimation using an embedded scheme~\citep{Hairer1996}. For BDF, the interpolation technique of Nordsieck is utilized~\citep{Nordsieck1962}. 

Because the final numerical solution error accumulates over the integration steps, it is important to choose tolerance levels wisely. A conservative rule of thumb is to choose tolerances that are two orders of magnitude smaller than the acceptable error. For a common INST \cmfa{} simulation, a relative tolerance of $tol_{rel}=10^{-6}$ and an absolute tolerance of $tol_{abs} = 10^{-9}$ mean that the IVP solution is about two orders of magnitude more accurate than the measurements $\mathcal{O}(10^{-4})$. These tolerance values are the default choice in \cfluxIII{}.

\subsection{Numerical solution quality}
\label{sec:solution-quality}
\subsubsection{Numerical accuracy for an analytically tractable test system}
\label{subsec:test-equation}

To test correctness of the IVP solver implementations in \cfluxIII{}, as it is common practice in the field, we test the accuracy of the numerical solution with a simple test equation
\begin{equation}
    \begin{pmatrix}
        A & 0 \\
        0 & 1
    \end{pmatrix} \cdot
    \frac{d}{dt} \begin{pmatrix}
        a \\
        b
    \end{pmatrix}
    =
    \begin{pmatrix}
        -(1 + \tau) && \tau \\
        1 + \tau && -1 \\
    \end{pmatrix}
    \cdot
    \begin{pmatrix}
        a \\
        b
    \end{pmatrix}
    +
    \begin{pmatrix}
        1 \\
        0
    \end{pmatrix}
    \label{eqn:test-ode}
\end{equation}
Despite its simplicity, this system exhibits the dynamic characteristics of isotope labeling systems: Increasing the value of $\tau$ and decreasing the value of $A$ worsens the condition of the system, making it increasingly stiff, and therefore more difficult to solve numerically. This is a well-known challenge for IVP solvers~\citep{Dahlquist1963}. 

The exact solution of Eq.~\eqref{eqn:test-ode} is given by
\begin{align}
    a_{\mbox{exact}} \left( t \right) = &\frac {1}{\left( 2\tau^2 - 2 \right)  \sqrt{{A}^{2}+ \left( 4\,{\tau}^{2}+2\,\tau-2 \right) A+ \left( 1+\tau \right) ^{2}} } \nonumber \\
    &\cdot \Biggl( \biggl(  2\,{\tau}^{2}+A+\tau+ \sqrt {{A}^{2}+ \left( 4\,{\tau}^{2}+2\,\tau-2 \right) A+ \left( 1+\tau \right) ^{2}} - 1 \biggr) \nonumber \\
    &\quad\quad \cdot {{\rm e}^{-\frac{ A+\tau+1-\sqrt {{A}^{2}+ \left( 4\,{\tau}^{2}+2\,\tau-2 \right) A+ \left( 1+\tau \right) ^{2}}}{2A} t}} \nonumber \\
    &\quad + \biggl(  -2\,{\tau}^{2}-A-\tau+ \sqrt {{A}^{2}+\left( 4\,{\tau}^{2}+2\,\tau-2 \right) A+ \left( 1+\tau \right) ^{2}} + 1\biggr) \nonumber \\
    &\quad\quad \cdot {{\rm e}^{-\frac { A+\tau+1+\sqrt {{A}^{2}+ \left( 4\,{\tau}^{2}+2\,\tau-2 \right) A+\left( 1+\tau \right) ^{2}} t}{2A} t}} \nonumber \\ 
    &\quad - 2 \sqrt{{A}^{2}+ \left( 4\,{\tau}^{2}+2\,\tau-2 \right) A+ \left( 1+\tau \right) ^{2}} \Biggr) \\
    \nonumber \\
    b_{\mbox{exact}} \left( t \right) = &\frac {1}{\left( 2\tau - 2 \right)  \sqrt{{A}^{2}+ \left( 4\,{\tau}^{2}+2\,\tau-2 \right) A+ \left( 1+\tau \right) ^{2}} } \nonumber \\
    &\cdot \Biggl( \biggl(  A+\tau+ \sqrt {{A}^{2}+ \left( 4\,{\tau}^{2}+2\,\tau-2 \right) A+ \left( 1+\tau \right) ^{2}} + 1 \biggr) \nonumber \\
    &\quad\quad \cdot {{\rm e}^{-\frac{ A+\tau+1-\sqrt {{A}^{2}+ \left( 4\,{\tau}^{2}+2\,\tau-2 \right) A+ \left( 1+\tau \right) ^{2}}}{2A} t}} \nonumber \\
    &\quad + \biggl(  -A-\tau+ \sqrt {{A}^{2}+\left( 4\,{\tau}^{2}+2\,\tau-2 \right) A+ \left( 1+\tau \right) ^{2}} - 1\biggr) \nonumber \\
    &\quad\quad \cdot {{\rm e}^{-\frac { A+\tau+1+\sqrt {{A}^{2}+ \left( 4\,{\tau}^{2}+2\,\tau-2 \right) A+\left( 1+\tau \right) ^{2}} t}{2A} t}} \nonumber \\ 
    &\quad - 2 \sqrt{{A}^{2}+ \left( 4\,{\tau}^{2}+2\,\tau-2 \right) A+ \left( 1+\tau \right) ^{2}} \Biggr)
    \label{eqn:test-ode-solution}
\end{align}

Knowing the exact solution allows defining the global numerical error of the numerical solution taken at $N$ time steps $t_i$ within the integration window $[0, T]$. Precisely, we define the global numerical error as follows
\begin{equation}
    e_{\textit{num,exact}} =
    \max\limits_{1 \le i \le N}
    \lVert \mathbf{x}_{\textit{num}}(t_i) - \mathbf{x}_{\textit{exact}}(t_i) \rVert_2, \ t_i \in [0, T]
    \label{eqn:sup-norm}
\end{equation}
Eq.~\eqref{eqn:sup-norm} represents the worst-case deviation of the calculated numerical solution from the analytical one across the entire integration steps. We use this error metric here to analyze the accuracy of the numerical IVP solvers that have been implemented in \cfluxIII.

With the test IVPs in Eq.~\eqref{eqn:test-ode} at hand, different parameter combinations of $\tau \in [10, 1000]$ and $A\in \left[1/10, 1/1000\right]$ are chosen to create test systems with different stiffnesses, as indicated by condition numbers ranging from low ($2 \cdot 10^2$) to high ($2 \cdot 10^6$) values. The IVPs are solved within the integration domain $[0, 100]$ with settings for low, medium, and high relative tolerances, i.e., $tol_{rel} = 10^{-3}, 10^{-6}$, and $10^{-9}$, respectively. In all cases, the absolute tolerance $tol_{abs}$ is set to $tol_{rel}\cdot 10^{-3}$ to prevent near-zero values from escaping the error control.

\begin{figure}[!ht]
    \centering
    \includegraphics[width=0.8\linewidth]{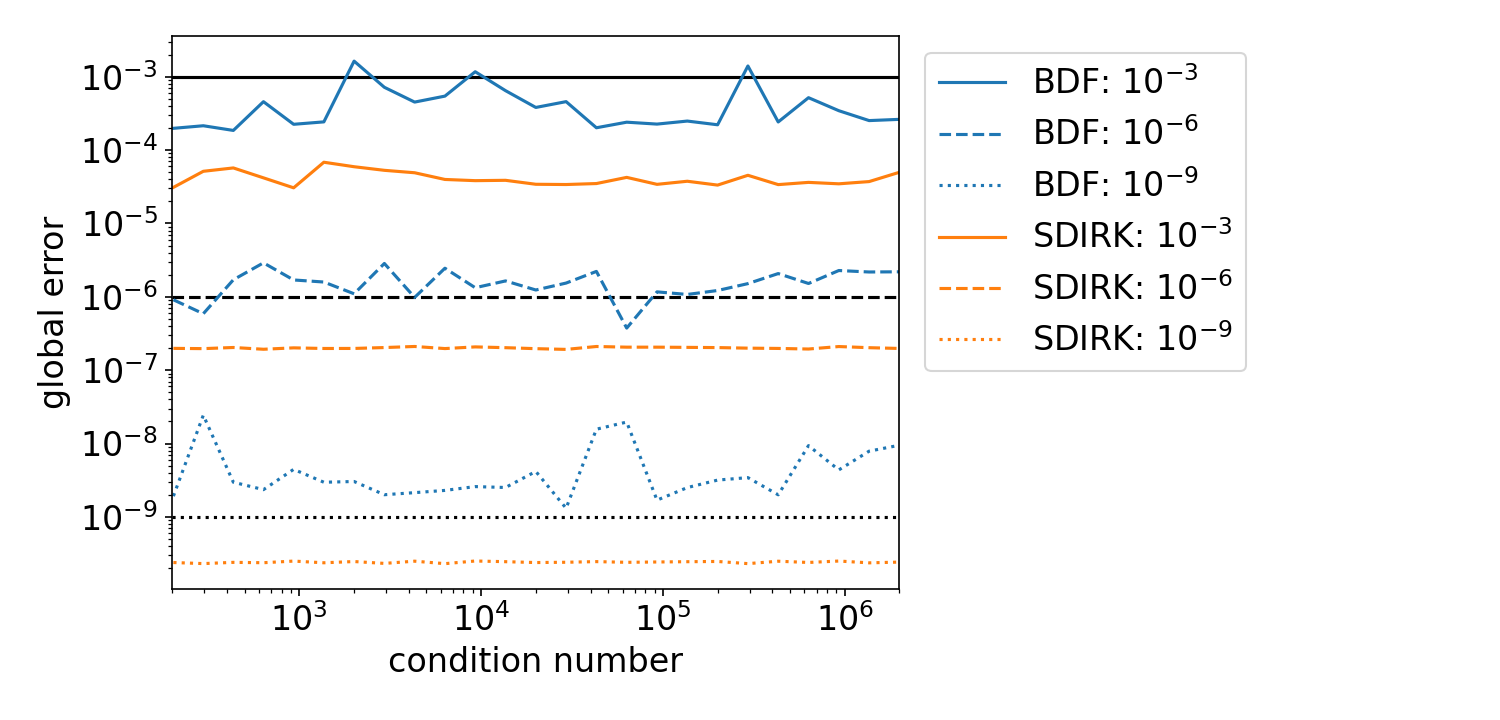}
    \caption{\textbf{Global numerical errors for test IVPs with differing degrees of ill-conditionedness}.
    The  BDF and SDIRK solvers implemented in \cfluxIII{} maintain a global numerical error $e_{\textit{num,exact}}$ close to the local tolerance, regardless of the condition number of the test IVP. The absolute tolerances are set to $tol_{rel}\cdot 10^{-3}$ in all cases.} 
    \label{fig:test-ode}
\end{figure}

Figure~\ref{fig:test-ode} shows the global numerical errors $e_{\textit{num,exact}}$ for the SDIRK and BDF solvers applied to the set of test IVPs. The achieved numerical accuracy is within one order of magnitude of the configured relative tolerance. Thus, both solvers show reliable control over the global numerical error and do not produce overly precise results, meaning that computational resources are not wasted by performing more time steps than necessary. However, for the test IVPs, we find that the step-size control of the SDIRK solver is more conservative than that of the BDF schemes. 
Abstracting from these results, we recommend selecting a relative tolerance of $10^{-6}$ for the BDF solver to achieve $\mathcal{O}(10^{-5})$ solution accuracy (see also SI~\Cref{subsec:ode-reliability}).

\subsubsection{Numerical reliability of INST simulations}
\label{subsec:ode-reliability}

For real-world \cmfa{} models, the exact solution is no longer analytically tractable. To assess the global numerical error for real \cmfa{} models, we therefore resort to a highly accurate SDIRK solution computed with a very small tolerance ($tol_{rel} = 10^{-12}$, $tol_{abs} = 10^{-15}$), which we use as reference IVP solution $\mathbf{x}_{ref}$. 
The degree of ill-conditionedness of the IVPs affect the accuracy of the numerical solutions. Therefore we randomly select 10 feasible parameter constellations and approximate the global numerical error $e_{num,ref}$ with respect to the reference solution according to 
\begin{equation}
    e_{\textit{num,ref}} =
    \max\limits_{1 \le i \le N}
    \lVert \mathbf{x}_{\textit{num}}(t_i) - \mathbf{x}_{\textit{ref}}(t_i) \rVert_2, \ t_i \in [0, T]
    \label{eqn:sup-norm2}
\end{equation}

\begin{figure}[!b]
    \centering
    \includegraphics[width=0.5\linewidth]{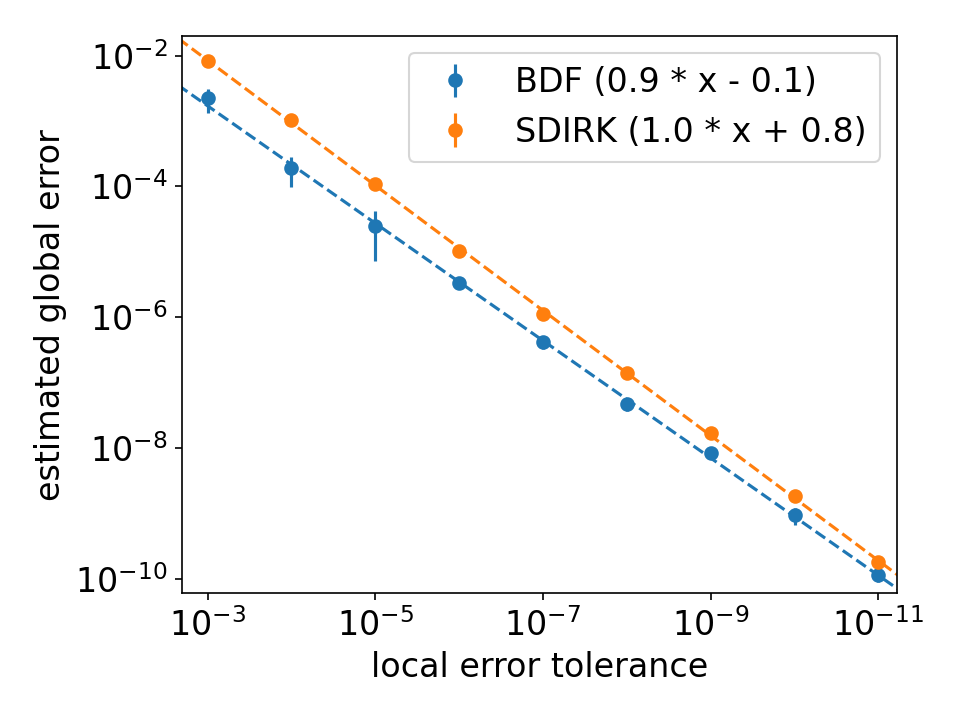}
    \caption{
    \textbf{Approximate global numerical error of SDIRK and BDF solvers for a real-world \cmfa{} model.}
    Mean approximated global numerical error $e_{\textit{num,ref}}$ for the model \Syn{\_a} and common ($10^{-3}-10^{-6}$) and high-accuracy tolerances ($10^{-6} - 10^{-11}$). Bars indicate standard deviations over random 10 parameter sets. The SDIRK reference solution is calculated with $tol_{rel} = 10^{-12}$ and $tol_{abs} = 10^{-15}$.}
    \label{fig:precision-scalability}
\end{figure}

For the \Synfull{} model \Syn{\_a} (see SI~\Cref{subsec:synechocystis} for details), \Cref{fig:precision-scalability} shows the mean and standard deviation of the approximated global numerical error for SDIRK and BDF solvers. We observe that the approximated global numerical error $e_{\textit{num,ref}}$ is linearly correlated (in log-log space) with the relative tolerance $tol_{rel}$. The desired tolerances are always met within one order of magnitude. Hence, we conclude that both numerical IVP solvers provide reliable and robust solutions to real-world \cmfa{} models. In particular, users of \cfluxIII{} have the ability to provide effective error control over the implemented numerical IVP solvers to meet individually preferences.

\subsubsection{Consistency of IST and INST numerical solutions}
\label{sec:inst_to_stat}

So far, we have evaluated the numerical accuracy of our INST solvers with different tolerances and realistic networks by comparing them to a ultrahigh-accuracy reference solution generated using the SDIRK solver implemented in \cfluxIII. Because there is no provably correct alternative simulator is available for independently verifying the numerical solution, we perform an additional consistency check utilizing the characteristics of the isotope labeling systems. Specifically, we verify that the labeling states of the INST IVPs in Eq.~\eqref{eqn:cascade-inst} converge to the solution of the IST equations in Eq.~\eqref{eqn:cascade-stat} for very large $t$ ~\citep{WiechertWurzel2001}.

Exploiting this fact allows us to study the consistency of the numerical IST and INST solutions, given a single FluxML model file. We here take the \ECfull{} \EC{\_a} model and calculate (i) the IST solution ($\mathbf{x}(\infty)$) and (ii) the labeling state at $t=10,000,000$~s  ($\mathbf{x}(10,000,000)$) for 1,000 random pool size parameter configurations. For each of these configurations, we determine the maximal absolute difference between the simulated IST and INST solutions in terms of the observed labeling patterns as follows

\begin{equation}
    e_{\textit{IST,INST}} = \lVert\mathbf{x} (\infty) - \mathbf{x}(10,000,000)\rVert_\infty
    \label{eqn:errorcheck}
\end{equation}

\begin{figure}[htb]
    \centering
    \includegraphics[width=0.5\linewidth]{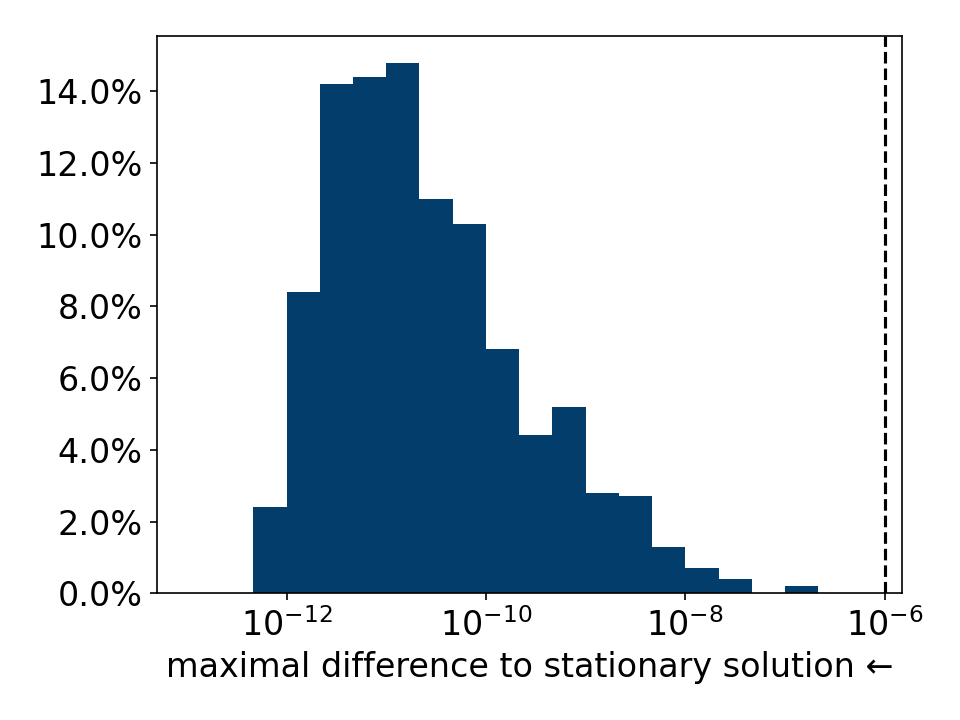}
    \caption{\textbf{\cfluxIII{} provides consistent IST and INST solutions.} 
    Worst-case differences between numerical IST and INST solutions, measured according to Eq.~\eqref{eqn:errorcheck} for the \EC{\_a} model and 1,000 random parameter sets. INST IVPs are solved using the BDF solver with $tol_{rel} = 10^{-6}$ (indicated by the vertical dashed line) and $tol_{abs} = 10^{-9}$.}
    \label{fig:instvstat}
\end{figure}

Figure~\ref{fig:instvstat} shows that the INST solution at a large $t$ generated using the BDF solver with $tol_{rel}= 10^{-6}$ and $tol_{abs} = 10^{-9}$ matches the IST solution with a maximum deviation of $1.2\cdot10^{-7}$, which is one order of magnitude lower than the solvers' relative tolerance. We therefore conclude that the numerical solution of the labeling systems in \cfluxIII{} is consistent and remains numerically accurate and reliable for very large integration time windows.

\subsection{Scalability of \cfluxIII{}}
\label{sec:sim_scalability}

In this section we investigate how the simulation runtime scales with increasingly complex measurement configurations and multiple datasets.
All runtimes are computed with a single core of an AMD EPYC 9334 CPU.

\subsubsection{Simulation of cumomer- and EMU-based labeling systems in \cfluxIII{}}
\label{subsec:labeling_scaling}

The simulation runtimes of the IST and INST labeling systems in Eq.~\eqref{eqn:cascade-stat} and Eq.~\eqref{eqn:cascade-inst} depend on the selected state space representations and their essential (i.e., dimension-reduced) dimensions (see Table~\ref{tab:dimensionalities} in SI~\Cref{subsec:labeling_dimensions}).
Exemplarily, here we benchmark the runtimes of the simulations of \cfluxIII{} IST and INST of the essential cumomer and EMU representations for the \ECfull{} model \EC{} (see SI~\Cref{subsec:ecoli} for details). For this, we use two configurations, a mass spectrometry (MS) from literature (\EC{\_a}) and a representative tandem mass spectrometry (MS/MS) configuration. For each of these two measurement configurations, we create 34 systems of essential customer and EMU dimensions through measurement subsampling. For each of these systems, we benchmark the simulation runtime for 1,000 randomly generated parameter sets.

\begin{figure}[!ht]
    \centering
    \includegraphics[width=0.85\linewidth]{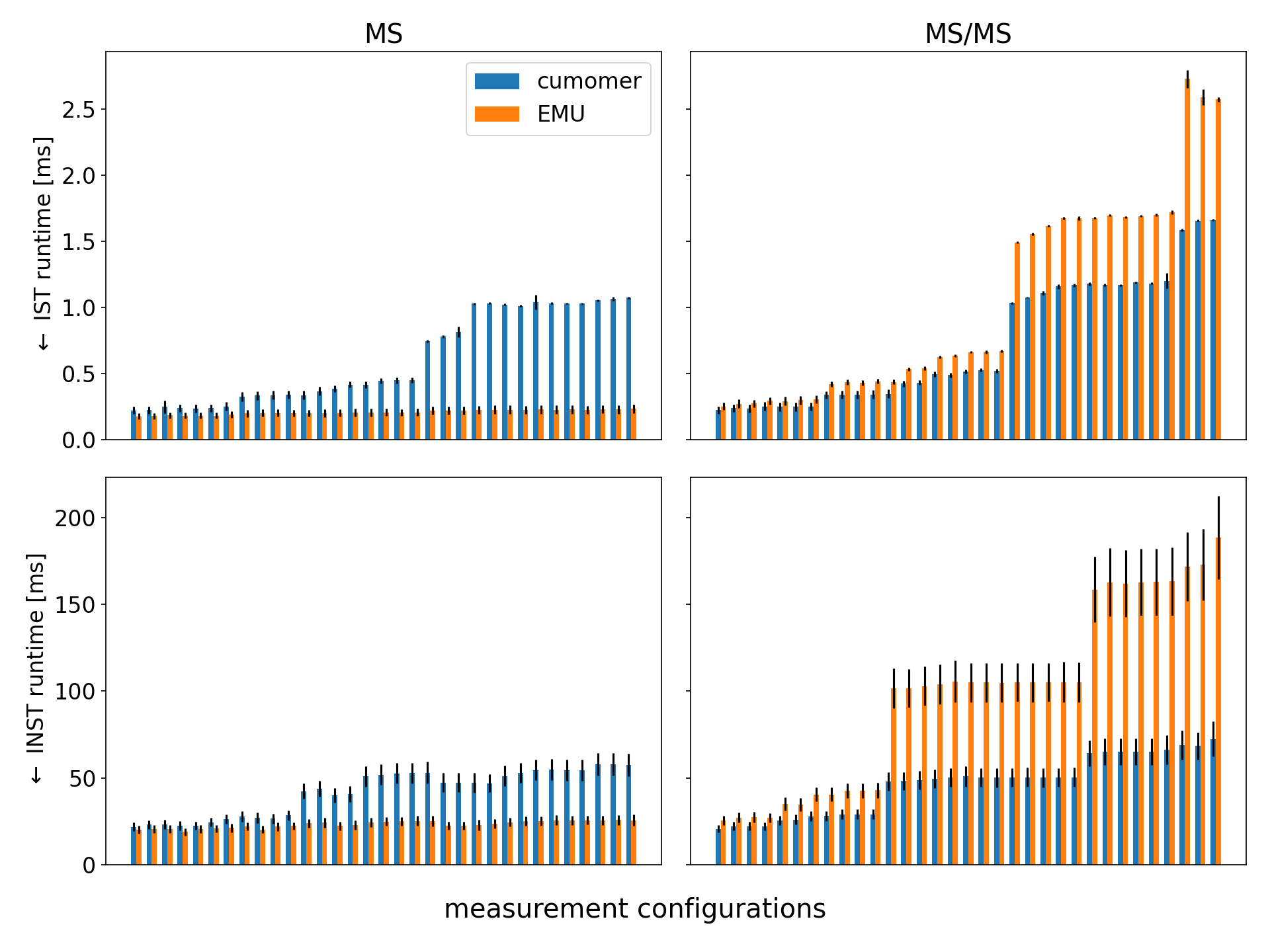}
    \caption{\textbf{Whether cumomer- or EMU-based labeling systems are faster to simulate depends on the measurement configuration.} Mean IST and INST simulation times for the \EC{} model and various MS and MS/MS measurement configurations for 1,000 parameter sets each. Bars indicate the standard deviation. IST (upper row) and INST (lower row) mean runtimes in milliseconds for essential cumomer (blue) and EMU (orange) labeling state representations.}
    \label{fig:runtimes_MS}
\end{figure}

\begin{figure}[H]
    \centering
    \includegraphics[width=0.75\linewidth]{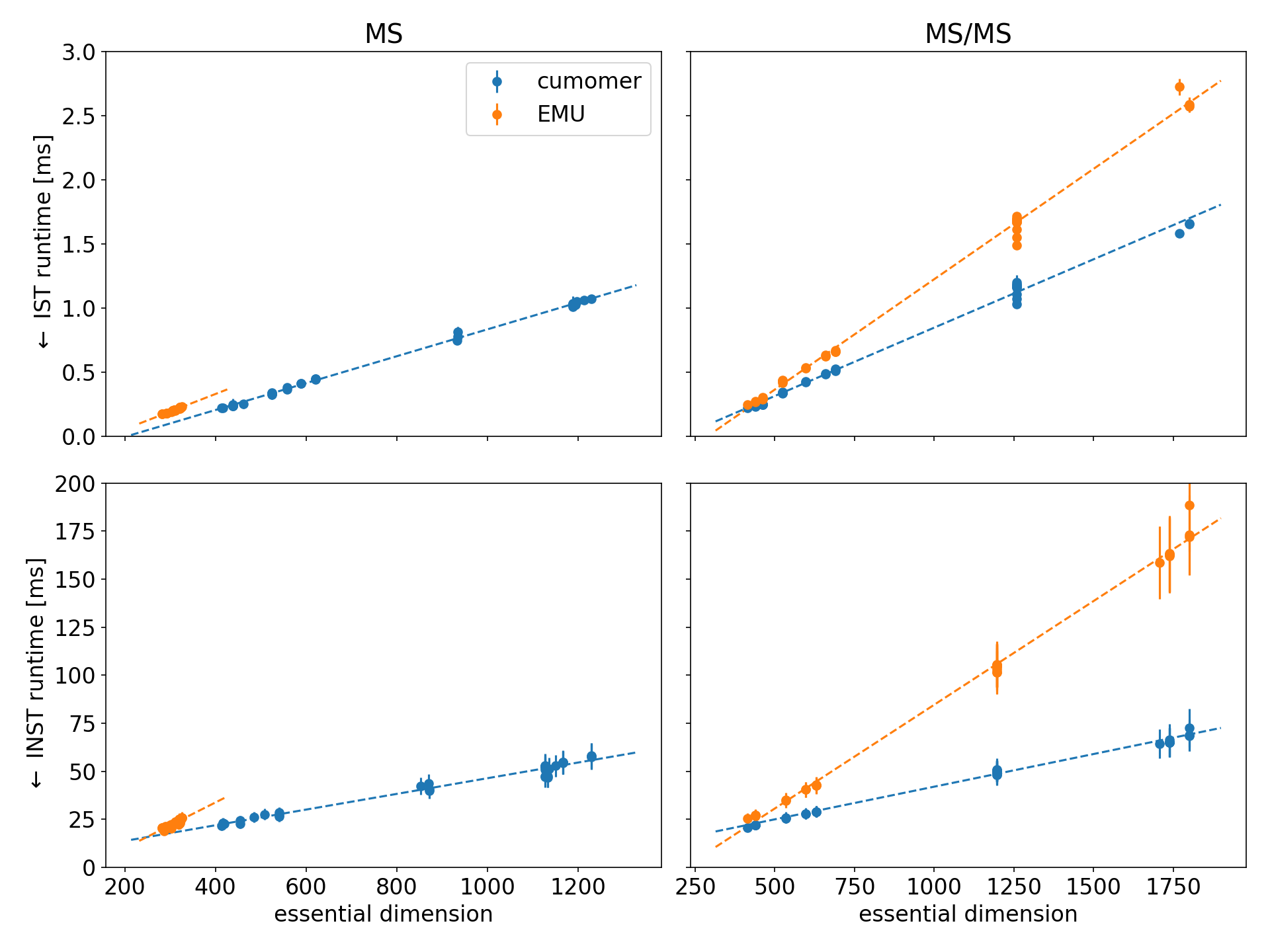}
    \caption{\textbf{Cumomer-based simulations are generally faster than EMU-based simulations when considering the same essential system dimensions.} Mean IST and INST simulation times for the \EC{} model with different MS and MS/MS configurations and random 1,000 parameter sets each. Bars indicate the standard deviation in simulation time.
    The relationship between essential cumomer/EMU labeling system dimensions and simulation times is highly correlated (Pearson coefficients $> 0.99$), pointing to \cfluxIII{}'s perfect scaling behavior. The scaling behavior is specific to the type of measurement configuration.}
    \label{fig:states}
\end{figure}

Figure~\ref{fig:runtimes_MS} shows a bar chart of the essential cumomer and EMU simulation times for the subsampled MS and MS/MS configurations. For MS measurement configurations, EMUs are significantly faster to simulate than cumomers for both, IST and INST labeling systems. This is due to the superior dimension-reduction capability of the essential EMU state-space formulation~\citep{Weitzel2007}. However, cumomer-based labeling systems are faster to simulate for MS/MS configurations, despite achieving similar state-space dimension reduction factors as EMU-based systems. This is because the vector-shaped form of cumomer systems is cheaper to simulate than the matrix-shaped EMU systems (see also \Cref{fig:states}). 
These results underscore the importance of choosing the most efficient labeling state representation for a given measurement configuration to achieve fast simulations, as implemented in \cfluxIII.

In \Cref{fig:states}, we plot the simulation times for the subsampled MS and MS/MS configurations over their the essential cumomer and EMU dimensions. We observe an almost perfect correlation for both IST and INST labeling systems, with Pearson coefficients of $>0.99$ in each case.
As expected, solving the EMU labeling systems is more costly than solving the cumomer labeling systems for the same essential system dimensionality. This is because cumomer-based labeling states are scalars, whereas EMU-based labeling states are vectors, and the underlying labeling systems have more unknowns to solve for.

Eventually, the most beneficial state-space representation is determined by the specific analytical method (MS - EMU, MS/MS - cumomer), while the scaling behavior within a given analytical measurement setting depends on the particular combination of labeling system structure and measurement configuration.

\subsubsection{Efficient simulation of datasets from multiple labeling experiments}
\label{subsec:multiple_ILEs_for_ED}

The swift computation of parameter sensitivities as a solution of the IVPs in Eqns.~\eqref{eqn:sensitivities-inst} and \eqref{eqn:sensitivities-stat} is crucial for all steps of the \cmfa{} workflow where Jacobian matrices are required. One example is the Fisher information matrix (FIM) from which the parameter covariance matrix is derived. Both matrices are central to experimental design (ED)~\citep{Beyß2021}.
The classic approach to isotope tracer ED is to explore the information content of all possible combinations of selected isotope tracers, which are commonly calculated sequentially~\citep{Mollney1999}. 
It is straightforward to see that, using a similar approach as described in SI~\Cref{subsec:multiple_ILEs}, we can reformulate the sequential ED calculation resembling Eq.~\eqref{eqn:mile-cascade-reformulated}. The reformulation enables to parallelize ED evaluations in \cfluxIII, either as a single one-shot evaluation or as a multiple-shot approach, split up into several evaluation batches with an adjustable size.
 
\begin{figure}[!ht]
    \centering
    \includegraphics[width=0.5\linewidth]{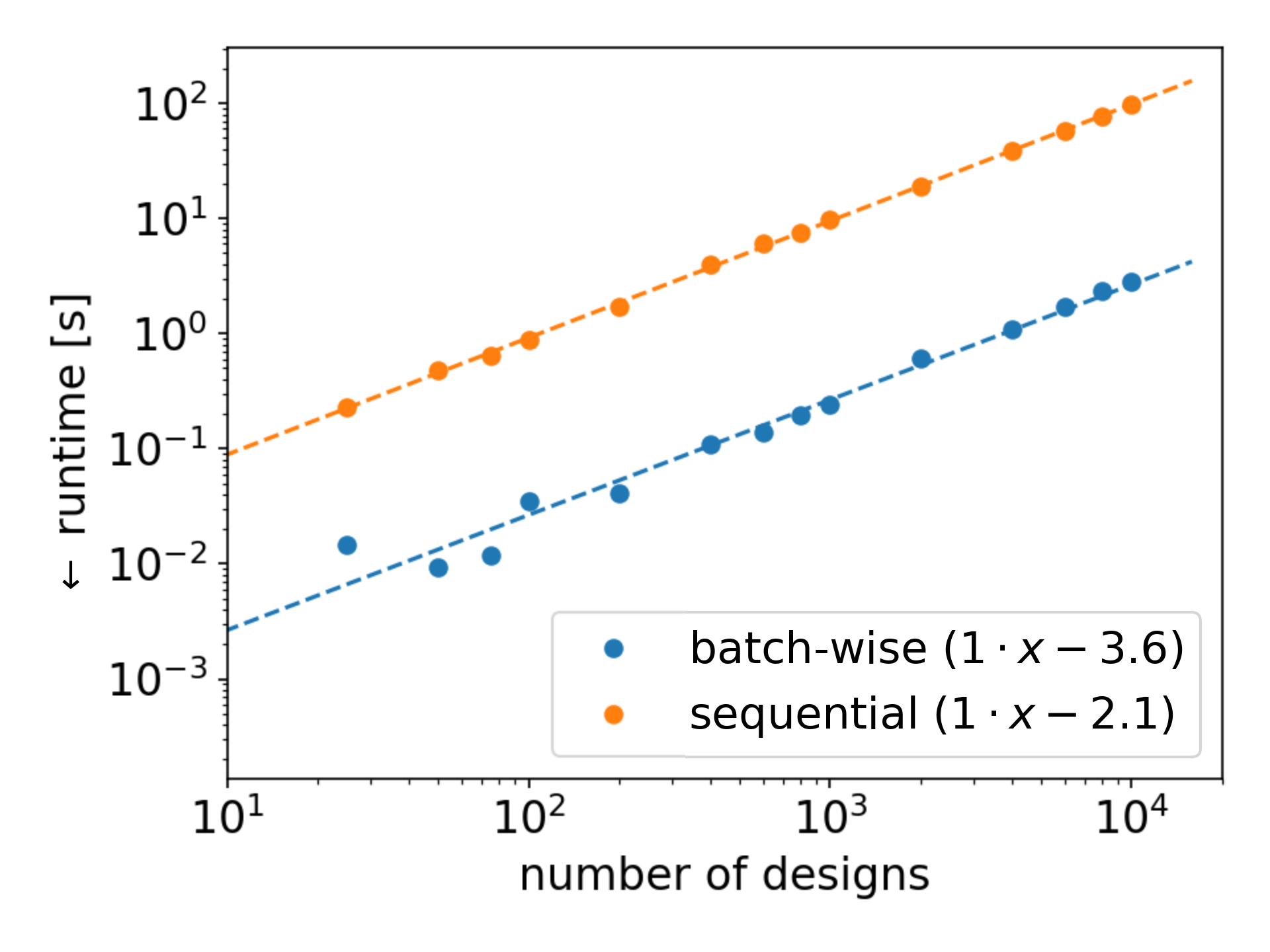}
    \caption{\textbf{Batch-wise ED simulation of labeling systems is one order of magnitude faster than the traditional sequential approach.} Comparison of IST sensitivity system runtimes for sequential and batch-wise approaches. For the \ECfull{} model \EC{\_a} the batch-wise approach scales consistently better than computing sensitivities sequentially.}
    \label{fig:ed-benchmark}
\end{figure}

For the \EC{\_a} model (see SI~\Cref{subsec:ecoli}), we compare the runtimes of the new batch-wise evaluation approach with those of the traditional sequential evaluation in Eq.~\eqref{eqn:mile-cascade}). The results in ~\Cref{fig:ed-benchmark} show that employing the batch-wise formulation reduces runtimes by a factor of about 20 (batch size = 75).
Consequently, using the batch-wise approach in \cfluxIII{} to calculate EDs allows more tracers to be considered or a finer grid to be explored within a given computational budget, compared to the conventional sequential approach. This improvement unlocks large-scale ED calculations.
\clearpage

\clearpage
\section{ \cfluxIII{} implementation details}
\label{sec:implementation_details}
\subsection{Software statistics}
\label{subsec:software_statistics}

The \cfluxIII{} code consists of 132 files, most of which are written in C++ and Python. \cfluxIII{} is built using \texttt{CMake} (\url{https://cmake.org/download/}). The breakdown of the lines of code (LOC) taken for a snapshot of the repository is given in Table~\ref{tab:loc}. Numbers are generated with \texttt{cloc . --exclude-dir=docs,extern,.readthedocs.yaml --exclude-ext=fml,xml,md,xslt} to not include documentation and examples.

\begin{table}[h]
    \centering
    \begin{tabular}{lrr}
    \toprule
         Language &  LOCs & files\\
         \midrule
         C++        & 14,062 & 108\\
         Python     &  1,763 &  10\\
         CMake      &    314 &   7\\
         YAML/TOML  &    272 &   4\\
         Dockerfile &     90 &   3\\
         \bottomrule
    \end{tabular}
    \vspace{0.2cm}
    \caption{Number of code lines of \cfluxIII{} resolved by programming language, quantified using \texttt{cloc}~\citep{Danial2021}. The numbers refer to versioned code at commit \textsf{43e0ef34f9a432f061b9790c52b085be4aa8ec5b} of the software repository.}
    \label{tab:loc}
\end{table}

\subsection{Interoperation of \cfluxIII{} with third-party libraries}
\label{subsec:interoperability}

Interoperability is a key quality of the \cfluxIII{} software. It allows the software to work seamlessly with other complementary tools, extending its core features. \cfluxIII{} communicates with third-party libraries via its carefully designed application programming interface (API), which provides a standardized interface for exchanging information bidirectionally. 
For example, the simulator object generated for a specific model is accessible and usable by other software tools. Also, embedding functionalities of third party tools into \cfluxIII{} is possible. We here demonstrate the latter for the case of numerical optimizers, while an example for the former is given in SI~\Cref{subsec:hopsy}.

Numerical optimization computes best-fit parameters in \cmfa{} by solving a linearly constrained nonlinear least-squares regression problem~\citep{Wiechert2015}. 
These regression problems exhibit a wide range of difficulties attributed to the ill-conditionedness of the inverse problem \citep{Theorell2017, Hadamard1902, Motulsky1987}. Thus, there is no one-size-fits-all solution that handles all optimization tasks equally efficiently. 
Imbuilt in \cfluxIII{} is the open gradient-based numerical optimizer \texttt{IPOPT} (Interior Point Optimizer)~\citep{Wachter2006}, taken from the COIN-OR project (\url{https://www.coin-or.org/})~\citep{Lougee-Heimer2003}, to solve the linearly constrained nonlinear least-squares regression problem. 
However, when a good starting point is lacking or the parameter domain is to be explored, a gradient-free optimizer may be a better choice.
In such situations, \cfluxIII{} makes it easy to use community optimizers and benefit from existing specialized algorithms and the continuous development of new, tailored ones.
Here, we demonstrate how easy it is to replace the built-in \texttt{IPOPT} optimizer with a third-party one. In this example, we use the standard \texttt{SciPy} optimizer \texttt{scipy.optimize.minimize}~\citep{Virtanen2020}.

\vspace{1em}
\small
\footnotesize
\begin{minted}{python}
import scipy.optimize as optimize
ineq_sys = simulator.parameter_space.inequality_system
result = optimize.minimize(simulator.compute_loss, starting_point, 
                           jac=simulator.compute_loss_gradient,
                           constraints=[optimize.LinearConstraint(ineq_sys.matrix, 
                                                                  ub=ineq_sys.bound, 
                                                                  keep_feasible=True)],
                           method="trust-constr")
\end{minted}
\normalsize
\vfill
\subsection{Expressive error messages}
\label{subsec:errors}
Coding errors are unavoidable when creating \cmfa{} evaluation workflows, particularly when new models or third-party software is involved (as described in SI~\Cref{subsec:interoperability}). Error messages and warnings are crucial to reducing the time spent on troubleshooting. However, detecting these errors and providing instructive solutions requires an extensive and careful analysis of possible reasons for those errors originating from technical (syntactic), contextual (logical), computational (numerical) and modeling (semantic) reasons. Therefore, we argue that providing expressive error messages is an important feature that distinguishes \cfluxIII{} from other flux analysis tools.

\cfluxIII{} provides more than 123 of such expressive messages (version 3.0.0a3). The following Table lists some examples.

\begin{flushleft}
\begin{tabular}
{p{0.12\linewidth} p{0.4\linewidth} p{0.4\linewidth}}
    \toprule
    \textbf{Error type} 
    & \textbf{Case} 
    & \textbf{Message} \\
    \midrule\addlinespace[10pt]
    \textit{semantic} 
    & Due to dimension reduction, all labeling states of a metabolite are eliminated from the cascaded labeling systems in Eq.~\eqref{eqn:cascade-inst}. In this case, the metabolite's pool size is to be eliminated from the IVP system, because it as no effect on the simulation outcome. This means that the pool size is under no circumstances identifiable from the given measurement setup and solving the respective part of the sensitivity system in Eq.~\eqref{eqn:sensitivities-inst} is superfluous.
    & \textit{Print warning} | Pool size of metabolite ``A'' is non-determinable given the measurement configuration, as it does not impact the simulation outcome. Consider fixing the pool size value in FluxML to reduce the ill-conditionedness of the inverse problem. \\\addlinespace[10pt]
    \textit{syntactic} 
    & Attempt to load a non-existing FluxML file. 
    & \textit{Throws ParseError} | Error with XML input source. \\\addlinespace[10pt]
    \textit{logic} 
    & Attempt to call labeling simulation or residual computation with infeasible model parameters, i.e., parameters that do not fulfill mass balances and/or violate inequality constraints. 
    & \textit{Throws ParameterError} | Free model parameters violate inequality constraints. Relax lower/upper bounds. If violations are very small, consider raising ``parameter{\_}space.constraint{\_}violation{\_}tolerance''. \\\addlinespace[10pt]
    \textit{numerical} 
    & Attempt to solve the INST IVP in Eq.~\eqref{eqn:cascade-inst} exceeds the number of maximal integration time steps, e.g. when very low error tolerances are requested.
    & \textit{Throws MathError} | Operation ``BDF (SUNDIALS) IVP solver'' failed: Maximum number of steps (here 100,000) reached. Increase maximum number of solver time steps or relax solver tolerances. \\
    \bottomrule
\end{tabular}
\end{flushleft}

\clearpage
\section{ Simulator collection}
\label{sec:simulators}

\begin{figure}[!ht]
    \includegraphics[width=0.92\textwidth]{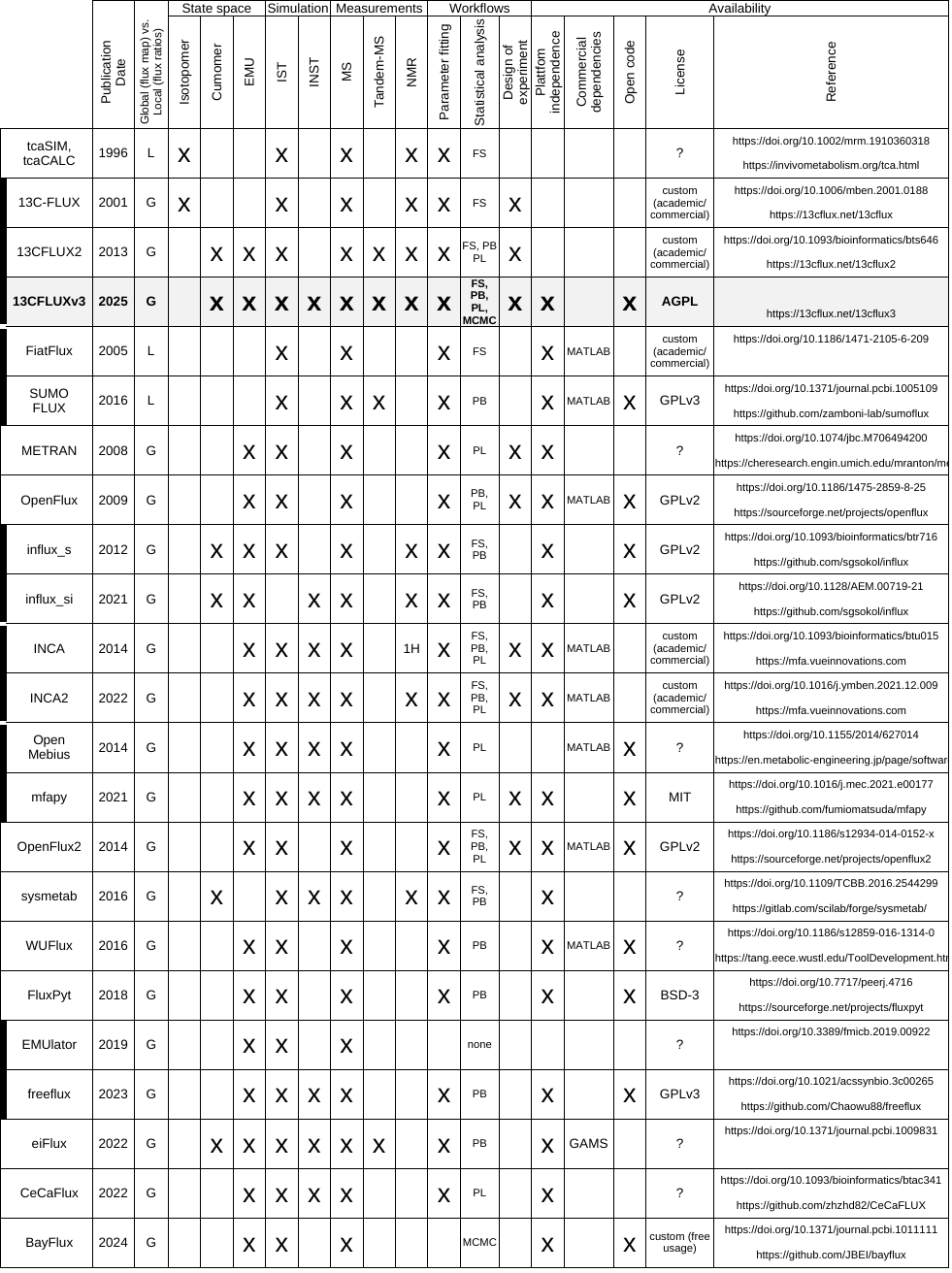}
    \captionof{table}{Common \cmfa{} simulators, chronologically ordered by their first published version. Multiple versions are depicted as a group, indicated by black bars on the left side. FS: Fisherian statistics, PB: parametric bootstrap, PL: profile likelihoods, MCMC: Markov chain Monte Carlo, for details see~\citep{Theorell2017}.}
\end{figure}

\clearpage
\section{ Simulator comparison}
\label{sec:simulator_comparison}

Next, we compare the simulation outcomes of \cfluxIII{} with those of three recent \cmfa{} simulators capable of performing IST and INST simulations, namely \texttt{INCA}, \texttt{FreeFlux}, and \texttt{influx\_SI}. All three simulators rely on EMUs as labeling state space representation. Besides numerical accuracy, we report simulation runtime benchmarks for the four simulators. Finally, the INST performance (accuracy versus runtime) scaling behavior of \cfluxIII{} is compared with that of \texttt{INCA}.
Because the simulators use slightly different natural \textsuperscript{13}C abundances, these are set to 0.010816 for all simulators. All benchmarks are executed on a single core Intel Core i7-4790 CPU.

\subsection{Comparison of simulated labeling states}
\label{subsec:label_comparison}

For the simulator comparison we use again the \ECfull{} model (see SI~\Cref{subsec:ecoli}), with its original MS measurement configuration (\EC{\_a}). We set the model parameters to the flux values $\mathbf{v}$ reported in the \texttt{INCA} v2.3 example \texttt{demo/ecoli/ecoli.m}, which we hitherto denote \textit{nominal fluxes}. We simulate the IST variant of the \EC{\_a} model and calculate the difference between the simulated labeling data of the three simulators -- \texttt{freeflux}, \texttt{INCA}, and \texttt{influx\_si} -- and that of \cfluxIII according to
\begin{equation}
    e_{y, \ast,\cfluxIII} ( \mathbf{v} )=
    \left| y_{\ast} (\mathbf{v}) - y_{\cfluxIII}(\mathbf{v}) \right|
    \label{eqn:diff_sim_stat}
\end{equation}
where $y$ denotes a single simulated measurement and the subscript $\ast$ represents one of the three simulators. 

\begin{figure}[!ht]
    \begin{tikzpicture}
        \node[anchor=north west, inner sep=0] (a) at(0, 0){\includegraphics[width=0.5\linewidth]{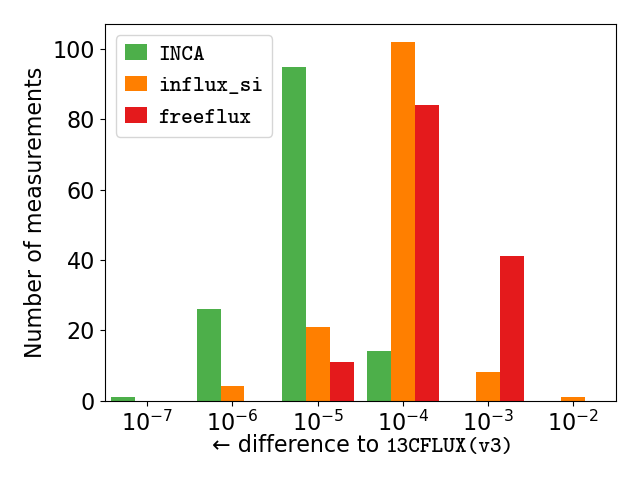}};
        \node[anchor=north west, inner sep=0] (b) at(0.5\linewidth, 0){\includegraphics[width=0.5\linewidth]{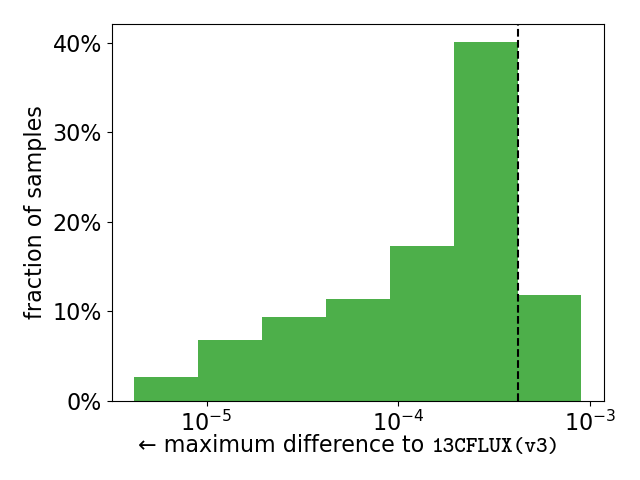}};
        \node[anchor=north west] at (0,0.3) {\large (a)};
        \node[anchor=north west] at (0.5\linewidth,0.3) {\large (b)};
    \end{tikzpicture}
    \caption{\textbf{Simulation differences between SOTA \cmfa{} tools for the \EC{\_a} model}. 
    (a) \textbf{IST \cmfa}. \cfluxIII{} and \texttt{INCA} show good agreement (\textless4.2e-4 worst-case deviation); differences between \cfluxIII{} and \texttt{influx\_si} are one order of magnitude larger (\textless4.4e-3). \texttt{freeflux} shows differences to the \cfluxIII{} simulation of up to 0.011, which is in the order of the measurement error.
    (b) \textbf{INST \cmfa.} Results for \cfluxIII{} and \texttt{INCA} are in good agreement. For almost 90\% of the samples the maximum difference is smaller than that of the IST case (a). The maximum absolute difference is $7.4\cdot 10^{-4}$. \cfluxIII{} uses the BDF solver with $tol_{rel}=10^{-6}$ and $tol_{rel}=10^{-9}$.}
    \label{fig:result_comp}
\end{figure}

Absolute differences $e_{y, \ast,\cfluxIII{}}$ from Eq.~\eqref{eqn:diff_sim_stat} are shown in \Cref{fig:result_comp}a. The numerical solutions of \cfluxIII{} and \texttt{INCA} differ by no more than $4.2 \cdot 10^{-4}$ (on average, $1.2 \cdot 10^{-4}$). We attribute this difference to the conversion between the different flux coordinate systems used by the simulators (\texttt{INCA} uses the forward/backward flux coordinate system, \cfluxIII{} the free net/exchange flux coordinate system). Nonetheless, the maximum difference in the solutions is an order of magnitude below the typical measurement standard deviation of $4 \cdot 10^{-3}$ and can therefore be considered negligible.
On the other hand, compared to the results of \cfluxIII{} and \texttt{INCA}, \texttt{freeflux} and \texttt{influx\_si} show significantly larger deviations of up to $1.1 \cdot 10^{-2}$ (average $2.7 \cdot 10^{-3}$) and $5.8 \cdot 10^{-2}$ (average $2.0 \cdot 10^{-3}$), respectively.

For the INST case, solving the labeling system is more sophisticated, and solution performances depend on flux and pool size values (see SI~\Cref{subsec:test-equation}). Systems of varying difficulty are generated by augmenting the nominal fluxes with 1,000 random pool size configurations ($\mathbf{X}_j$ same as in SI~\Cref{sec:inst_to_stat}). Due to limitations of \texttt{freeflux} and \texttt{influx\_si} in terms of solution performance (see below), we are required to restrict our  comparison to \cfluxIII{} and \texttt{INCA}.

For every parameter set we compute the maximum absolute difference between the simulated labeling data vectors $\mathbf{y}(t)$ produced by the two simulators at the observed time points $t_i$ in the set of measurement time points
\begin{equation}
    e_{\texttt{INCA}, \cfluxIII} ( \mathbf{v}, \mathbf{X}_j) = \max_{1\leq i \leq N}\  \lVert \mathbf{y}_{\cfluxIII} (t_i, \mathbf{v}, \mathbf{X}_j) - \mathbf{y}_{\texttt{INCA}} (t_i, \mathbf{v}, \mathbf{X}_j) \rVert_\infty,\  j = 1(1)1,000
    \label{eqn:x3_inca}
\end{equation}
Both tools use the EMU state-space representation with solver tolerances set to $tol_{rel}=10^{-6}$ and $tol_{abs} = 10^{-9}$.

The distribution of differences between the outcomes of the two simulators, $e_{\texttt{INCA}, \cfluxIII}$ from Eq.~\eqref{eqn:x3_inca}, is shown in \Cref{fig:result_comp}b.  
For about 90\% of the parameter sets, the difference in terms of $e_{\texttt{INCA}, \cfluxIII{}}$ is smaller than the largest IST difference, $e_{y, \texttt{INCA},\cfluxIII}$, while the maximal absolute difference in simulated labeling states across the parameter set is $7.4 \cdot 10^{-4}$.  Consequently the differences are in the same order of magnitude as in the IST case. 

In conclusion, \texttt{INCA} and \cfluxIII{} deliver comparable simulated labeling states for IST and INST with small differences that are likely to be attributed to their use of different flux coordinate systems.
The differences between \cfluxIII{} and \texttt{freeflux}, as well as \texttt{influx\_si}, are larger in the IST case, whereas they cannot be properly assessed in the INST case.

\subsection{Performance benchmark}
\label{subsec:benchmark}

To compare the time to perform forward simulation using \cfluxIII{} to that of \texttt{freeflux}, \texttt{INCA} and \texttt{influx\_si}, we set up a benchmark with the \EC{} (IST and INST) and \Syn{} (INST) models (see SI~\Cref{subsec:labeling_dimensions} for details). 
Simulations with two different labeling measurement configurations are performed because these configurations significantly impact the system dimension (see \Cref{tab:dimensionalities}). 
In addition, the INST run times are taken for three IVP solver tolerances, namely $tol_{rel} = 10^{-3}$ (low accuracy), $tol_{rel} = 10^{-6}$ (standard accuracy), and $tol_{rel} = 10^{-9}$ (high accuracy). In all cases the absolute tolerance $tol_{abs}$ is set to $10^{-3}\cdot tol_{rel}$. 
The simulation task is repeated 100 times with the same parameter set. Mean and standard deviation are reported.
Results are summarized in \Cref{tab:times}. 

\begin{table}[!ht]
\centering
    \renewcommand{\arraystretch}{0.91} 
    \begin{tabular}{lllr@{\hspace{2pt}}c@{\hspace{2pt}}rlr@{\hspace{2pt}}c@{\hspace{2pt}}rlr@{\hspace{2pt}}c@{\hspace{2pt}}rlr@{\hspace{2pt}}c@{\hspace{2pt}}r}
    \toprule
                           & Model              & \makecell{accuracy\\($tol_{rel}$)} & \multicolumn{3}{c}{\cfluxIII{}$^{\dagger}$} &  & \multicolumn{3}{c}{\texttt{freeflux}\textsuperscript{*}} &  & \multicolumn{3}{c}{\texttt{INCA}} &  & \multicolumn{3}{c}{\texttt{influx\_si}\textsuperscript{**}} \\
    \midrule\addlinespace[10pt]
    \multirow{15}{*}{\rotatebox[origin=c]{90}{INST}} & \multirow{3}{*}{\EC{\_a}}  & $10^{-3}$     & 40.95       & ±     & 0.81      &  & 214.17     & ±    & 1.38     &  & 296     & ±  & 123 &  & 225.07     & ±     & 18.62     \\
                           &                        & $10^{-6}$     & 113.51      & ±     & 3.60      &  & \multicolumn{3}{c}{N/A}          &  & 2.790   & ±  & 145 &  & \multicolumn{3}{c}{N/A}           \\
                           &                        & $10^{-9}$     & 334.87      & ±     & 12.85     & & \multicolumn{3}{c}{N/A}          &  & 43,397  & ±  & 237 & & \multicolumn{3}{c}{N/A}           \\\addlinespace[10pt]
                           & \multirow{3}{*}{\EC{\_b}}  & $10^{-3}$     & 20.17       & ±     & 0.80      &  & 188.43     & ±    & 30.60    &  & 207     & ±  & 66  &  & 57.55      & ±     & 12.31     \\
                           &                        & $10^{-6}$     & 54.89       & ±     & 0.50      & & \multicolumn{3}{c}{N/A}          &  & 759     & ±  & 85  & & \multicolumn{3}{c}{N/A}           \\
                           &                        & $10^{-9}$     & 159.13      & ±     & 3.78      & & \multicolumn{3}{c}{N/A}          &  & 15,371  & ±  & 399 & & \multicolumn{3}{c}{N/A}           \\\addlinespace[10pt]
                           & \multirow{3}{*}{\Syn{\_a}} & $10^{-3}$     & 32.54       & ±     & 1.42      &  & 221.47     & ±    & 1.85     &  & 199     & ±  & 89  &  &            &       &           \\
                           &                        & $10^{-6}$     & 96.58       & ±     & 4.85      & & \multicolumn{3}{c}{N/A}          &  & 810     & ±  & 157 &  &            &       &           \\
                           &                        & $10^{-9}$     & 278.50      & ±     & 7.37      &  & \multicolumn{3}{c}{N/A}          &  & 35,937  & ±  & 388 &  &            &       &           \\\addlinespace[10pt]
                           & \multirow{3}{*}{\Syn{\_b}} & $10^{-3}$     & 24.46       & ±     & 0.38      &  & 161.94     & ±    & 0.73     &  & 147     & ±  & 47  &  &            &       &           \\
                           &                        & $10^{-6}$     & 67.87       & ±     & 2.77      & & \multicolumn{3}{c}{N/A}          &  & 548     & ±  & 62  &  &            &       &           \\
                           &                        & $10^{-9}$     & 201.35      & ±     & 5.80      & & \multicolumn{3}{c}{N/A}          &  & 16,359  & ±  & 315 &  &            &       &           \\\addlinespace[10pt]
    \multirow{2}{*}{\rotatebox[origin=c]{90}{IST}}  & \EC{\_a}                   &          & 0.45        & ±     & 0.01      &  & 15.25      & ±    & 0.08     &  &    0\textsuperscript{\#}        &    &        &  & 6.71       & ±     & 2.48      \\
                           & \EC{\_b}                   &          & 0.26        & ±     & 0.02      &  & 6.93       & ±    & 0.14     &  &     0\textsuperscript{\#}       &    &        &  & 2.58       & ±     & 0.40     \\\addlinespace[10pt]
    \bottomrule
    \end{tabular}
    \vspace{0.2cm}
    \caption{\textbf{Simulator performance comparison.} Reported simulation times in [ms] are the mean ± standard deviation of 100 runs.
    N/A indicates that the simulator is not able to compute the desired configuration or cannot be configured as indicated.
    \textsuperscript{$\dagger$}~EMU state-space representation.
    \textsuperscript{$\ast$}~\texttt{freeflux} has no possibility to influence the IVP solver tolerance. Therefore, runtimes are reported in the $10^{-3}$ accuracy row, although the achieved accuracy likely deviates. 
    \textsuperscript{$\ast\ast$}~The IVP solver of \texttt{influx\_si} has no automated step-size control. By manually adapting the step-size, a numerical accuracy of roughly $10^{-3}$ is achieved. When testing smaller step-sizes, the IVP solver generates numerically instable results.
    \textsuperscript{\#}~Due to limited runtime measurement resolution, IST runtimes are often reported to be 0. 
    }
    \label{tab:times}
\end{table}
\newpage

Clearly, INST simulations take longer than IST simulations. For INST, the solver tolerance significantly affects the runtime (see also Figure~1B in the main text). 
A runtime comparison of \EC{} and \Syn{} model variants reveals that, besides the network size (i.e., the number of metabolites and reactions) the effective dimensions of the dimension-reduced state spaces (see \Cref{tab:dimensionalities}) also play an important role.

The most important observations regarding the simulators are:
\begin{enumerate}
    \item \texttt{freeflux} does not provide access to the IVP solvers step-size, the solver tolerance is not adaptable and the delivered simulation accuracy remains elusive.
    \item \texttt{influx\_si} has no automated step-size control and the inbuilt setting fails to solve the system at higher accuracies.
    \item \cfluxIII{} outperforms \texttt{freeflux}, \texttt{INCA}, and \texttt{influx\_si} across all settings.
\end{enumerate}

\subsection{Parameter variation}
\label{subsubsec:param_variation}

\begin{figure}[!ht]
    \centering
    \includegraphics[width=1\linewidth]{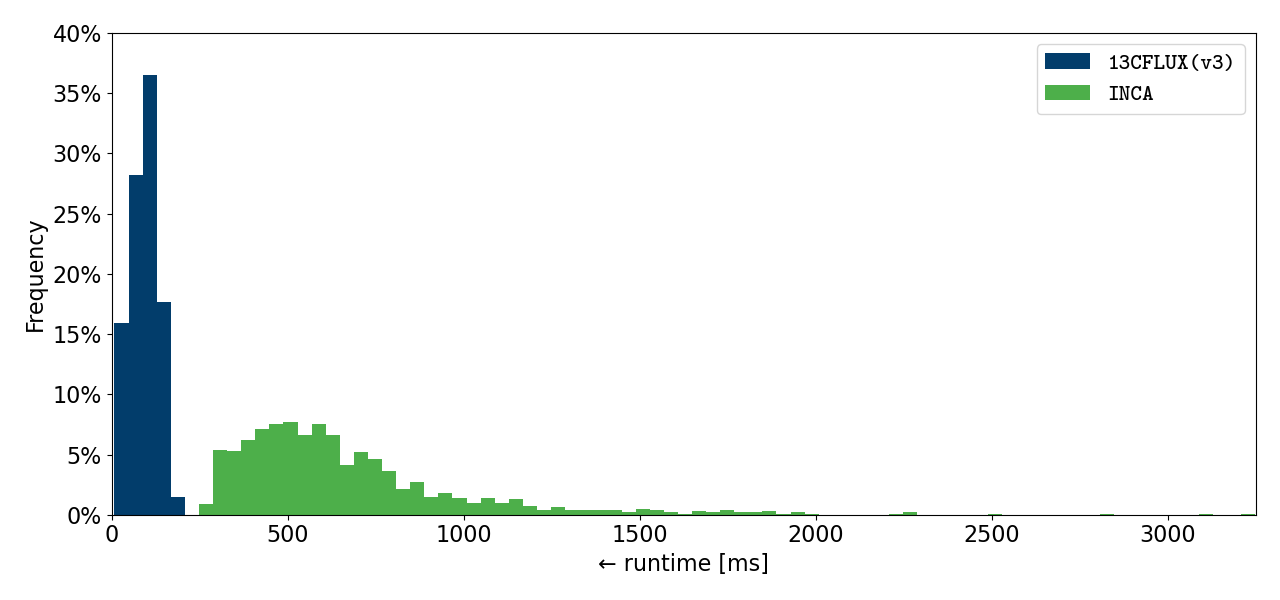}
    \caption{\textbf{INST runtime variation for different sets of pool sizes.} \cfluxIII{} shows a narrow distribution pointing to a very efficient step-size control.}
    \label{fig:poolsize_runtime}
\end{figure}

We quantify the effect of varying pool sizes on INST simulation times for the \textit{E.~coli} model \EC{\_a}. 1,000 random pool size parameter sets from \Cref{sec:inst_to_stat} are used for simulations with  both the \cfluxIII{} and the \texttt{INCA} simulators. The tolerances for both simulators are set to $tol_{rel} = 10^{-6}$ and $tol_{abs}=10^{-9}$. 
Each parameter set is calculated five times, and the mean value for the runtime recorded.
Mean runtime distributions are are shown in \Cref{fig:poolsize_runtime}.

For \cfluxIII, the runtime ranges from 15~ms to 207~ms (median 92.5~ms). For \texttt{INCA}, the median runtime is a factor of 6 larger (582~ms), and its distribution range is approximately 15 times larger (from 258~ms to 3,246~ms) than that of \cfluxIII{}.

In conclusion, \cfluxIII{} has shorter simulation times on average that are more robust with respect to parameter variations.
While the variation in IST simulation times over different flux sets is nearly negligible, the runtime for INST depends strongly on the model parameters. Therefore, when benchmarking INST runtimes different parameter sets must be considered.
\newpage
\subsection{Scalability of INST simulations}
\label{subsec:scalability}

\begin{figure}[!ht]
    \centering
    \includegraphics[width=0.5\linewidth]{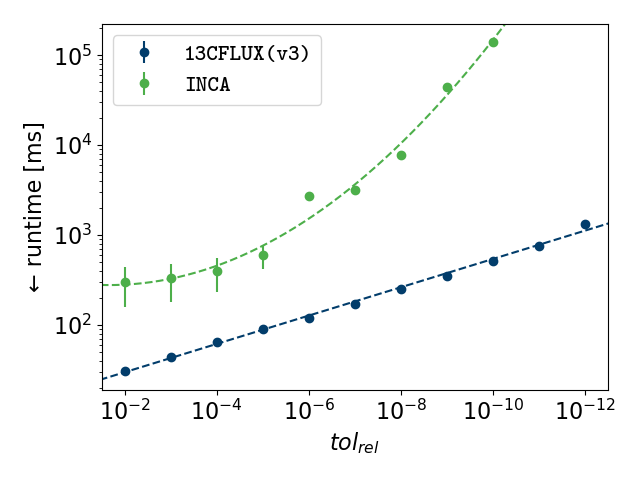}
    \caption{\textbf{\cfluxIII{} INST log runtimes scale linearly with decreasing log solver tolerances, whereas those of \texttt{INCA} scale quadratically.}
    Log-log plot of the mean runtimes (dots) taken for the \EC{\_a} model, with error bars originating from 100 repetitions. Linear and quadratic regression lines of $log_{10}(runtime)$ for \cfluxIII{} (blue: $-0.1575 \cdot log_{10}(tol_{rel}) + 1.1582$) and \texttt{INCA} (green: $0.0392\cdot \left(log_{10}(tol_{rel})\right)^2 + 0.1297 \cdot log_{10}(tol_{rel}) + 2.5504$), respectively.}
    \label{fig:acc_v_runtime}
\end{figure}

Finally, we compare the performance trade-off between simulation times and numerical accuracy for relative tolerances between $tol_{rel} = 10^{-2}$ and $10^{-12}$ (with $tol_{abs} = 10^{-3} \cdot tol_{rel}$).
\cfluxIII{} shows linear correspondence between the log-tolerances and the log-runtimes in \Cref{fig:acc_v_runtime}. 

The scaling behavior of \texttt{INCA} runtimes is comparable for relative tolerances of $10^{-2}$ and $10^{-5}$ (with an offset of one order of magnitude in runtimes). However, they grow log-quadratically with decreasing (log) tolerances.
For \cfluxIII{}, an increase in relative tolerance of one order of magnitude requires approximately 40\% more computing time. 

\clearpage
\section{ Novel application: Bayesian INST \cmfa{}}
\label{sec:bayesian-mfa}

The Bayesian \cmfa{} approach has several advantages over optimization-based parameter fitting~\citep{Theorell2024}, which still dominates the field. Most notably, the Bayesian approach provides a richer understanding of the probability of plausible parameter values and their dependencies through the \textit{posterior probability distribution}. Analyzing the posterior distribution enables the detection of non-identifiable parameters and the examination of parameter correlations, offering a holistic understanding of parameter uncertainty. This is particularly appealing for INST \cmfa{}, where the correlation between fluxes and pool sizes may be quite informative despite the fact that the single parameters may be not practically identifiable.

To estimate the posterior distribution of model parameters, their prior distribution, which captures existing knowledge, is combined with their likelihood in view of the data~\citep{Theorell2017}. Stochastic Markov chain Monte Carlo (MCMC) methods approximate the posterior distribution of the fluxes by constructing a Markov chain that generates thousands to millions of random samples (simulations)~\citep{Brooks2011}. These methods have strong theoretical underpinnings and asymptotic convergence guarantees to the true posterior under relatively mild conditions.

For IST \cmfa{}, the Bayesian approach to parameter estimation has been described several times; see, for instance, ~\citet{Kadirkamanathan2006, Theorell2017, Backman2023r, Hogg2023}. However, the Bayesian approach has not yet been attempted for INST \cmfa{}. One reason is that, for MCMC to converge, many samples must be drawn (in the order of $10^6-10^9$), making fast simulation crucial. On the other hand, the parameter space to be sampled has complex polytopic geometry, requiring the construction of tailored Markov chains that efficiently traverse these spaces~\citep{Jadebeck2023}.

To unlock Bayesian INST \cmfa{}, we combine \cfluxIII{} with the highly optimized polytope sampling tool \hopsy{}~\citep{Paul2024}. \texttt{hopsy} relies on the high-performance polytope sampling library \texttt{HOPS}, which is tailored toward efficiently exploring the complex parameter spaces underlying \cmfa{} models~\citep{Jadebeck2021}.

\subsection{Efficient MCMC sampling using \cfluxIII{}}
\label{subsec:hopsy}
\cfluxIII{} provides the function \mintinline{python}{run_non_uniform_sampling} to facilitate MCMC sampling of model parameters.  
To benefit from state-of-the-art MCMC algorithms, \cfluxIII{} leverages the MCMC sampling platform \hopsy.
High performance sampling analyses are maintained through the use of numpy arrays, the de facto standard for multidimensional data in Python, for communication.
The expressive API of \cfluxIII{} enables the integration with \hopsy{} in just a few lines of Python code:

\vspace{0.5em}
\small
\footnotesize
\begin{minted}[xleftmargin=20pt,linenos]{python} 
def run_non_uniform_sampling(simulator, num_samples: int, starting_point: np.ndarray,
                             bounds: Dict[str, Tuple[float, float]],
                             num_chains: int,
                             proposal: hopsy.PyProposal, 
                             random_seed: int,
                             **kwargs):
    problem = hopsy.Problem(ineq_constr_matrix, ineq_constr_bound, HopsyModel(simulator))
    if starting_point is None:
        problem.starting_point = hopsy.compute_chebyshev_center(problem)
    else:
        problem.starting_point = starting_point

    mcs, rngs = hopsy.setup(problem, random_seed, n_chains=num_chains, proposal=proposal)
    _, samples = hopsy.sample(mcs, rngs, n_samples=num_samples, n_procs=num_chains, **kwargs)

    return samples
\end{minted}
\normalsize
\vspace{0.5em}
\noindent In line 7, the \cfluxIII{} simulator object is encapsulated by a \mintinline{python}{x3cflux.HopsyModel} object, which is part of the \cfluxIII{} API. 
The \mintinline{python}{x3cflux.HopsyModel} object manages the translation from \cfluxIII{} (\texttt{compute\_loss}) to \hopsy{} (\texttt{log\_density}) according to
\vspace{1em}
\small
\footnotesize
\begin{minted}[xleftmargin=20pt,linenos]{python}
class HopsyModel:
    def __init__(self, simulator):
        self.simulator = simulator

    def log_density(self, x):
        return -0.5 * self.simulator.compute_loss(x)
\end{minted}
\normalsize
\vspace{1em}
\noindent Line 5-6 encodes the translation between the \cfluxIII{} (C++) and \hopsy{} (Python). 
Specifically, the code lines define the likelihood log-density in \hopsy{} as the negated and halved residuals computed by \cfluxIII{}.
This code exemplifies, how the functionality of \cfluxIII{} is extended to perform efficient MCMC-based statistical analyses.

\subsection{Application}
\label{subsec:BayesApplication}

We now perform Bayesian INST \cmfa{} using \cfluxIII, where we select the \Synfull{} model \Syn{\_a} as use case (see SI~\Cref{sec:models} for details). To be able to assess the quality of the posterior probabilities provided by Bayesian inference, we rely on a synthetic dataset, simulated for realistic parameters, which are perturbed according to typical measurement errors (\texttt{syn\_perturbed.fml}). For the pool sizes normal prior distributions are formulated, exchange fluxes are constraint by an upper bound of 100. Four Markov chains are then run from independent dispersed initial starting points. We employ parallel tempering~\citep{Geyer1991} to tackle potential multi-modalities of the posterior distribution.
The chains are run for 2,400,000 samples, which were reduced to 24,000 samples by using thinning (thinning factor of 100). 

The computational workflow is set up in the Python script (\texttt{S6\_1-INST-mcmc.py}). Since the Bayesian analysis of INST \cmfa{} is compute intense, we provide a Docker container containing the workflow script, \cfluxIII{} and \hopsy{}, which can be executed on a workstation or compute server. We recommend using Slurm (Slurm scripts are reproduced by \texttt{S6\_0-create-SLURM-scripts.py}) to parallelize the sampling workflow across nodes.
Alternatively, the samples can be reproduced on a single compute node using the command

\vspace{1em}
\small
\footnotesize
\begin{minted}{bash}
    docker run -v .:/task jugit-registry.fz-juelich.de/ibg-1/modsim/fluxomics/13cflux:latest \
               /task/run-sampling.sh
\end{minted}
\normalsize
\vspace{1em}
\noindent To check for the convergence of the (thinned) Markov chains, it is common to calculate the rank-normalized potential scale reduction factor (PSRF) or $\hat{R}$ value~\citep{Vehtari2021} and check whether its values for each parameter is below a threshold close to one.
As a final visual quality check, posterior predictive plots are generated by forward simulating a representative subset of the posterior samples.

\begin{figure}[!htp]
    \centering
    \includegraphics[width=0.9\linewidth]{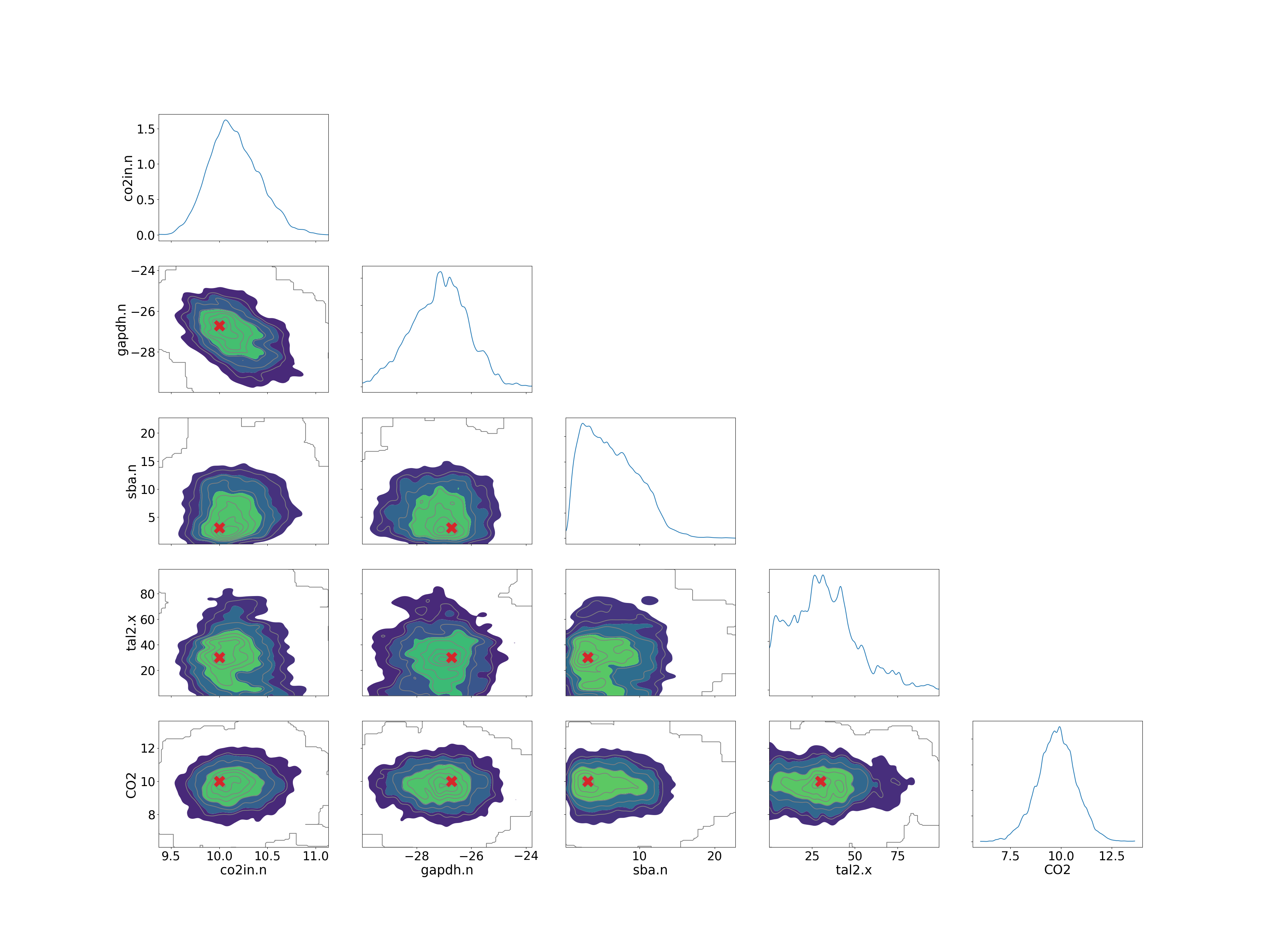}
    \caption{\textbf{Marginal and 2D INST \cmfa{} parameter posterior distributions for \Syn{\_a} model.} MCMC sample distributions are smoothed using kernel density estimation. 
    The red crosses mark the ground truth values. For each free parameter the rank-normalized potential scale reduction factor $\hat{R}$ is below $1.05$.}
    \label{fig:posterior}
\end{figure}

Figure~\ref{fig:posterior} shows marginal and joint posterior distributions for selected parameters (net fluxes and pool sizes). The red crosses mark the ground truth solution. The posterior probability modes match the ground truth parameters well. 
The results reveal a non-normality of the marginal posterior distribution of the \textit{sba.n} net flux, the fact that the \textit{tal2.x} exchange flux is nearly non-identifiable, and that the net fluxes \textit{co2in.n} and \textit{gapdh.n} are correlated. Furthermore, the marginal posterior for the \textsf{CO2} metabolite resembles its prior.

\begin{figure}[!htp]
    \centering
    \includegraphics[width=0.9\linewidth]{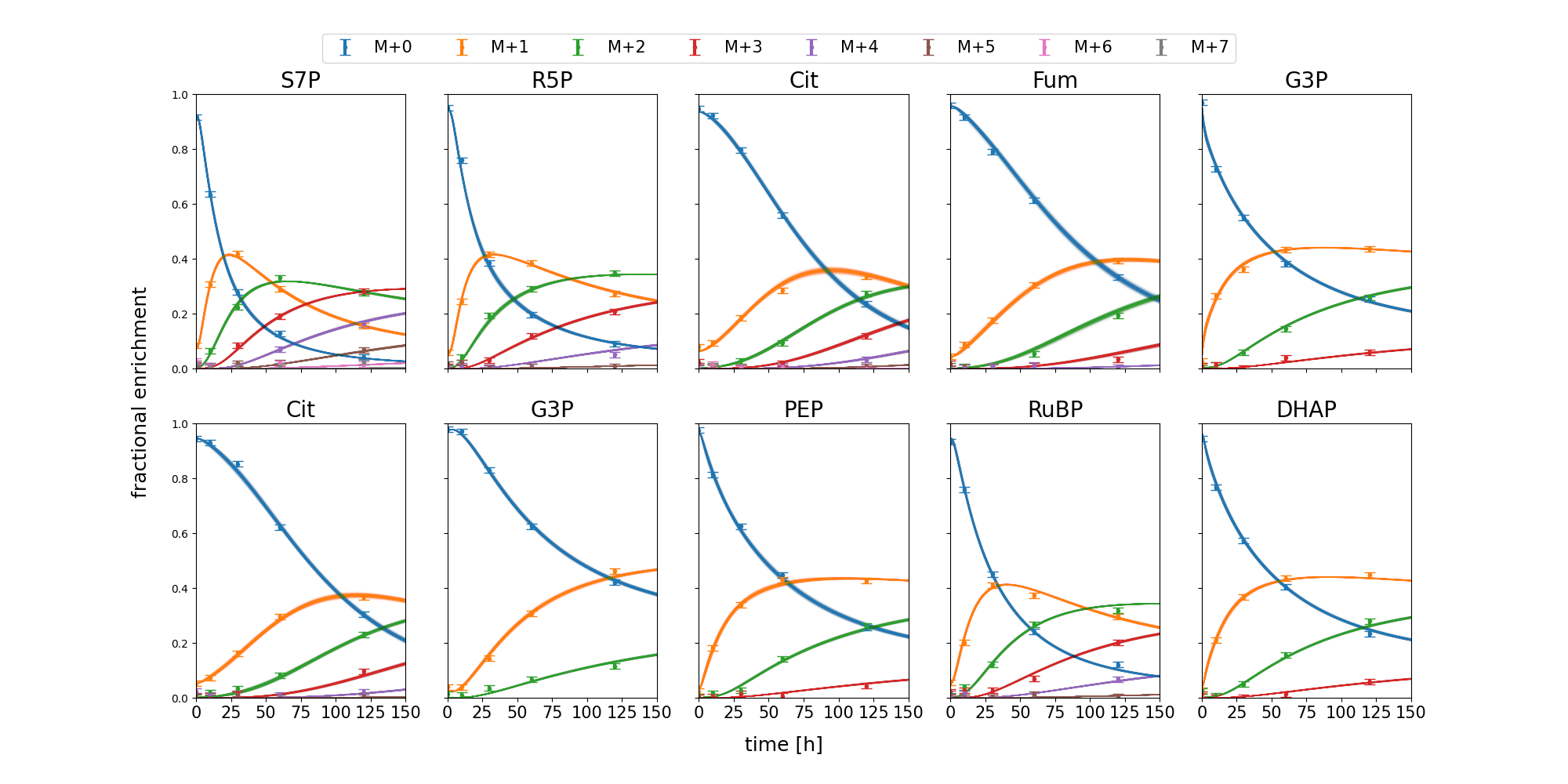}
    \caption{\textbf{Posterior predictive plots for the \Syn{\_a} model.} 100 samples are used to simulate data (lines) and plotted with the data points (capped bars).}
    \label{fig:predictives}
\end{figure}

Figure~\ref{fig:predictives} shows exemplary posterior predictive plots. The posterior predictive labeling curves are in excellent agreement with the data and exhibit a low prediction variance. This indicates that the model perfectly matches the data, as expected for the synthetic case studied here.

In summary, by its interoperability with the MCMC toolbox \hopsy{} and the portability enabled by Docker, \cfluxIII{} unlocks convenient and computationally scalable Bayesian INST \cmfa{} workflows.

\clearpage
\section{ \Air'ing \cfluxIII{} production workflows}
\label{sec:airflow}

Scientific workflows delineate the sequence of computational steps involved in a flux analysis, such as executing a simulation of a \cmfa{} network model or performing statistical inference on experimental datasets~\citep{Dalman2013}. When formalized as a directed graph -- in which nodes represent inputs and analysis tasks, and edges define the dependencies between them -- such workflows offer a precise structural representation of the analysis pipeline performed to handle the task at hand. This formalization promotes both the reproducibility and repeatability of evaluations, which improves the overall scientific quality. Moreover, the representation facilitates workflow automation for routine analyses, where the same steps are applied over and over again. This automation reduces manual efforts and minimizes potential errors.

Apache \Air{} is a powerful workflow orchestration platform that is actively maintained by a large user community (\href{https://airflow.apache.org/}{https://airflow.apache.org/}). Lately, \Air{} has also gained traction for MLops (machine learning operations), which involves deploying and maintaining machine learning models in production environments, proving an excellent choice for automating and monitoring software.

\begin{figure}[!ht]
    \centering
    \includegraphics[width=1.\linewidth]{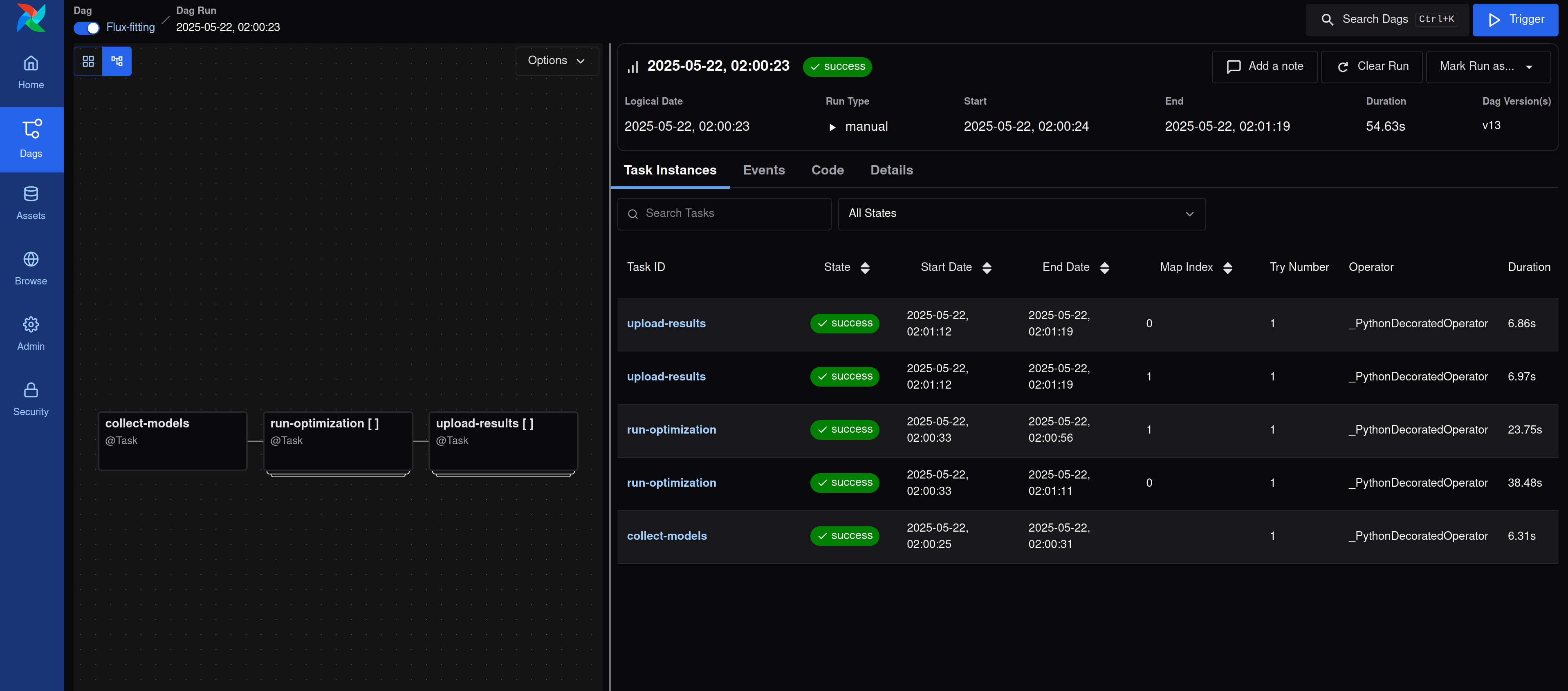}
    \caption{\textbf{\Air{}}DAG for flux fitting. \texttt{Airflow-3} provides a modern web user interface for managing and monitoring workflows. Two FluxML models are collected from the cloud storage, the dynamic task mapping creates two \textit{run-optimization} and two \textit{upload-results} tasks.}
    \label{fig:scientificworkflow}
\end{figure}

In \Air, workflows are defined using DAGs (directed acyclic graphs). DAGs are set-up in Python and automatically visualized in the \Air{} user interface (see \Cref{fig:scientificworkflow}). The nodes in a DAG represent tasks that need to be executed, and \Air{} automatically computes their dependencies based on the provided workflow constraints, such as task \textit{A} must be completed before task \textit{B}. Tasks are executed by \textit{executors}, which offer flexibility in terms of where and how to run a task. For example, computationally intensive tasks can be run on workstations, either natively or within  Docker images, providing a well-defined compute environment. Additionally, \Air{} provides monitoring and alerting capabilities, enabling process and status tracking of the workflow execution.

Figure~\ref{fig:scientificworkflow} shows an \Air{} DAG for a simple \cmfa{} fitting workflow, which is intentionally kept simple (excluding statistical tests, visualizations etc.). In the first workflow step, users upload their models to a S3 cloud storage device, in the so-called S3 bucket. DAGs are defined to run on a schedule or triggered manually, as in this case. Once the flux fitting DAG is triggered, it collects the models from the S3 bucket. Using dynamic task mapping, the tasks to be run are adjusted to match the number of flux models. For each model, a multi start optimization is run and the results as well as their associated model are uploaded to a separate S3 location, where they can be collected by users. Further useful tasks are e-mail notifications, informing the users as soon as the workflow is finished.
\clearpage
The Python listing for this \cmfa{} workflow is shown below. The code can be easily extended by adding steps such as statistical or visual quality controls. Note that, in addition to an \Air{} instance, an S3 server is required to execute this code.
\small
\footnotesize
\begin{minted}{python} 
    import boto3
    import datetime
    import logging
    import json
    import os
    import tempfile
    import x3cflux
    from pathlib import Path
    from airflow.sdk import DAG, task
    from airflow.models import Variable
    logger = logging.getLogger(__name__)
    
    s3_id = Variable.get("s3_id")
    s3_secret =  Variable.get("s3_secret")
    s3_endpoint = Variable.get("s3_endpoint")
    s3_bucket = "13CFLUX3"
    remote_prefix = "models/"
    results_prefix = "results/"
    
    with (DAG(
            dag_id="Flux-fitting",
            start_date=datetime.datetime(2025, 1, 1),
            schedule="@daily",
    )):
        @task(task_id="collect-models")
        def collect_models(id, secret, bucket, endpoint, remote_prefix=remote_prefix):
            session = boto3.session.Session()
            s3_client = session.client(
                service_name="s3",
                aws_access_key_id=id,
                aws_secret_access_key=secret,
                endpoint_url=endpoint,
            )
            response = s3_client.list_objects_v2(Bucket=bucket, Prefix=remote_prefix, Delimiter="/")
            fml_files = []
            for fml_file in response.get("Contents", []):
                if fml_file['Key'].endswith('.fml'):
                    fml_files.append(fml_file['Key'])
            return fml_files
    
    
        @task(task_id="run-optimization")
        def run_optimization(fml_file, id, secret, bucket, endpoint, remote_prefix=remote_prefix):
            session = boto3.session.Session()
            s3_client = session.client(
                service_name="s3",
                aws_access_key_id=id,
                aws_secret_access_key=secret,
                endpoint_url=endpoint,
            )
            with tempfile.TemporaryDirectory() as tmp:
                os.mkdir(os.path.join(tmp, remote_prefix))
                local_name = os.path.join(tmp, fml_file)
                with open(local_name, "wb") as f:
                    s3_client.download_fileobj(bucket, fml_file, f)
                simulator = x3cflux.create_simulator_from_fml(local_name)
                # Starts multi-fit from uniformly distributed points
                starting_points = x3cflux.run_uniform_sampling(simulator, 10_000)[0].T
                logger.info(f'running multistart optimization for {local_name}')
                optima, losses = x3cflux.run_multi_optimization(simulator, starting_points, num_procs=8)
                return {"fml_file": fml_file, "optima": optima.tolist(), "losses": losses.tolist()}
    
    
        @task(task_id="upload-results")
        def upload_results(optimization_result, id, secret, bucket, endpoint, remote_prefix=remote_prefix):
            session = boto3.session.Session()
            s3_client = session.client(
                service_name="s3",
                aws_access_key_id=id,
                aws_secret_access_key=secret,
                endpoint_url=endpoint,
            )
            fml_file = optimization_result['fml_file']
            model_name = Path(fml_file).stem
            json_name = f"{model_name}_optimization_result.json"
            with tempfile.TemporaryDirectory() as tmp:
                os.mkdir(os.path.join(tmp, remote_prefix))
                local_name = os.path.join(tmp, remote_prefix, json_name)
                with open(local_name, "w") as f:
                    json.dump(optimization_result, f)
                logger.info(f'uploading {local_name}')
                # Moves fml file from models
                s3_client.upload_file(local_name, bucket, os.path.join(results_prefix, json_name))
    
                local_fml_name = os.path.join(tmp, fml_file)
                with open(local_fml_name, "wb") as f:
                    s3_client.download_fileobj(bucket, fml_file, f)
                s3_client.upload_file(
                    local_fml_name,
                    bucket,
                    os.path.join(results_prefix, os.path.basename(fml_file))
                )
    
                delete_response = s3_client.delete_object(
                    Bucket=bucket,
                    Key=fml_file
                )
    
        # collects models to fit from s3
        fml_files = collect_models(id=s3_id, secret=s3_secret, bucket=s3_bucket, endpoint=s3_endpoint)
        #  runs optimization
        optimization_results = run_optimization.partial(
            id=s3_id, secret=s3_secret, bucket=s3_bucket, endpoint=s3_endpoint
        ).expand(fml_file=fml_files)
        #  uploads results
        upload_results.partial(
            id=s3_id, secret=s3_secret, bucket=s3_bucket, endpoint=s3_endpoint
        ).expand(optimization_result=optimization_results)
    \end{minted}
\normalsize
\clearpage

\section{ Metabolic models}
\label{sec:models}

In this work, \cmfa{} network models are utilized in different variations. In this section, most important facts about the base models are shortly summarized.

\subsection{Linear pathway model}
\label{subsec:linea}

\begin{figure}[!ht]
    \centering
    \includegraphics[width=0.5\linewidth]{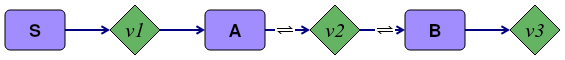}
    \caption{Model of a linear pathway with one exchange flux.}
    \label{fig:linea}
\end{figure}

\noindent  The linear pathway model represents a linear reaction chain with the external metabolite \textsf{S} (substrate) and two internal metabolites (\textsf{A}, \textsf{B}). \textsf{A} and \textsf{B} are connected via a bidirectional reaction \textit{v\textsubscript{2}} (see \Cref{fig:linea}). All metabolites carry one (carbon) atom. With the pool size $B=1$, the net fluxes \textit{v\textsubscript{1,net}} $=$  \textit{v\textsubscript{2,net}} $=$ \textit{v\textsubscript{3,net}} $ = 1$, the exchange flux \textit{v\textsubscript{2,xch}} $ = \tau$, and fully labeled input substrate, the labeling states of the metabolites \textsf{A} and \textsf{B} of this linear reaction chain coincide with the test ODE in SI~\Cref{subsec:test-equation}.

\noindent Profile of the linear pathway model:
\begin{description}
    \item[Reaction network:] 3 metabolites (2 intracellular), 3 reactions (2 intracellular, 1 bidirectional)
    \item[Independent parameters:] 1 net flux, 1 exchange flux, 2 pool sizes. We fix one pool size (\textsf(B)) and one net flux, thus only one unknown exchange flux and one unknown pool size remain.
    \item[Measurement configuration:] 1 mass spectrometry (MS) measurement of \texttt{B} at 1 time point \item[Labeling system: ] 4  isotopomer states, 2 cumomer or EMU states, $K = 1$ cascade levels
\end{description}
\clearpage

\subsection{\textit{Escherichia coli} (\EC{})}
\label{subsec:ecoli}
\begin{figure}[!ht]
    \centering
    \includegraphics[width=1\linewidth]{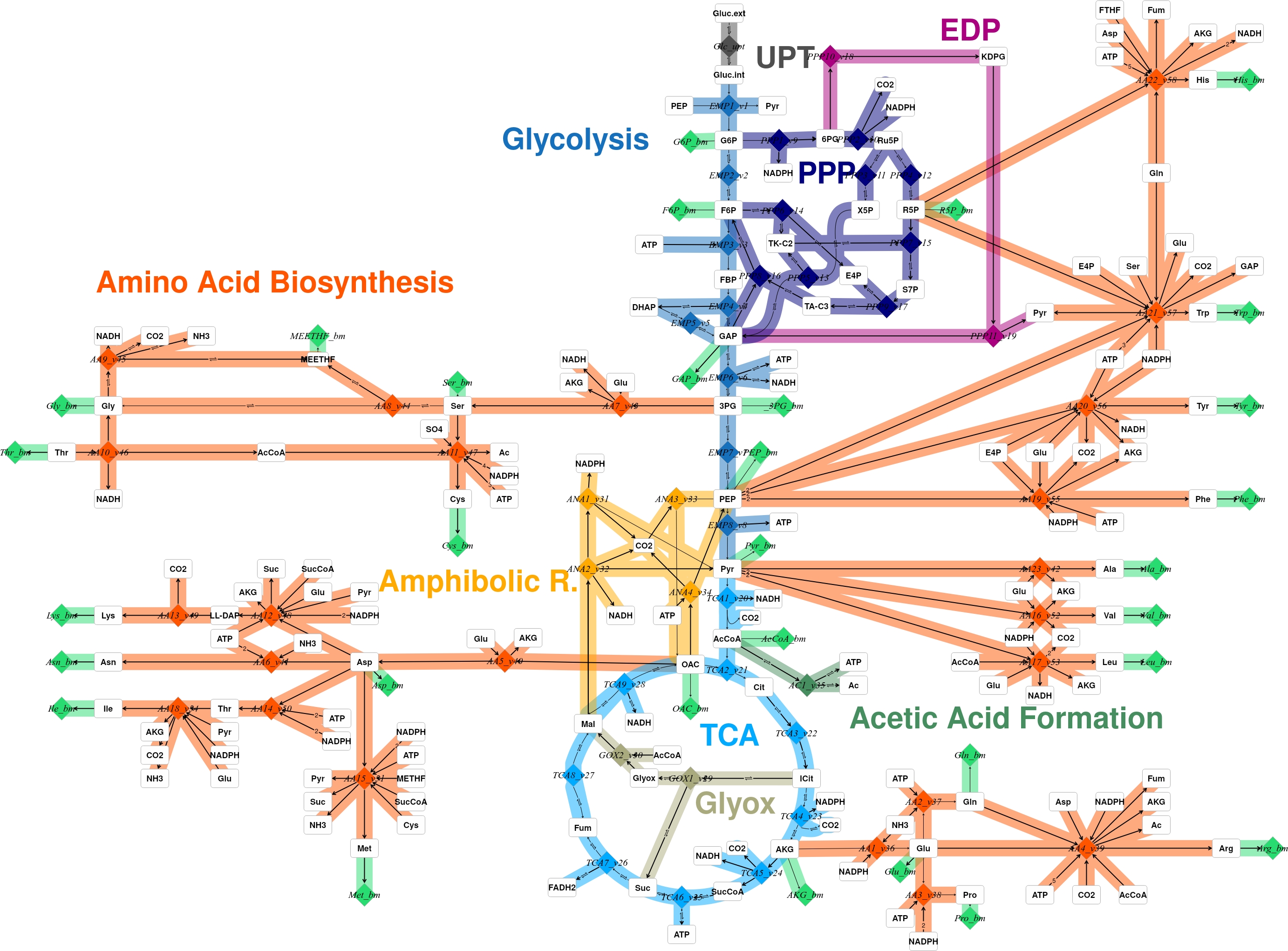}
    \caption{\textbf{Network model of \ECfull{}}. Throughout the text the model is referred to as \EC.}
    \label{fig:ec}
\end{figure}

\noindent  \EC{} is a model of the central carbon metabolism of \ECfull{} including a detailed formulation of amino acid formation. The model is provided with the \texttt{INCA} software version 2.3 (\texttt{demo/ecoli/ecoli.m} gives rise to a variant for INST \cmfa{} from which we have created an IST model variant). It comprises 100 reactions connecting 60 metabolites, features 4,858 isotopomers, and has 12/30/56 free model parameters (net fluxes/exchange fluxes/pool sizes). 
Two measurement configurations are available for both IST and INST \cmfa:
The original measurement configuration, termed \EC{\_a}, has 33 (partially doubled) MS measurements of amino acid fragments and intermediates, for IST taken at 1 and for INST taken at 9 equidistant measurement time points, resulting in a total of 120 and 1,161 independent labeling measurements for IST and INST, respectively.
A second measurement configuration, termed \EC{\_b}, has one MS measurement of an Alanine (Ala) fragment at the same measurement time points as \EC{\_a}, resulting in a total of 2 and 18 independent labeling measurements for IST and INST, respectively.

\clearpage
\subsection{\Synfull{} (\Syn)}
\label{subsec:synechocystis}

\begin{figure}[!ht]
    \centering
    \includegraphics[width=0.4\linewidth]{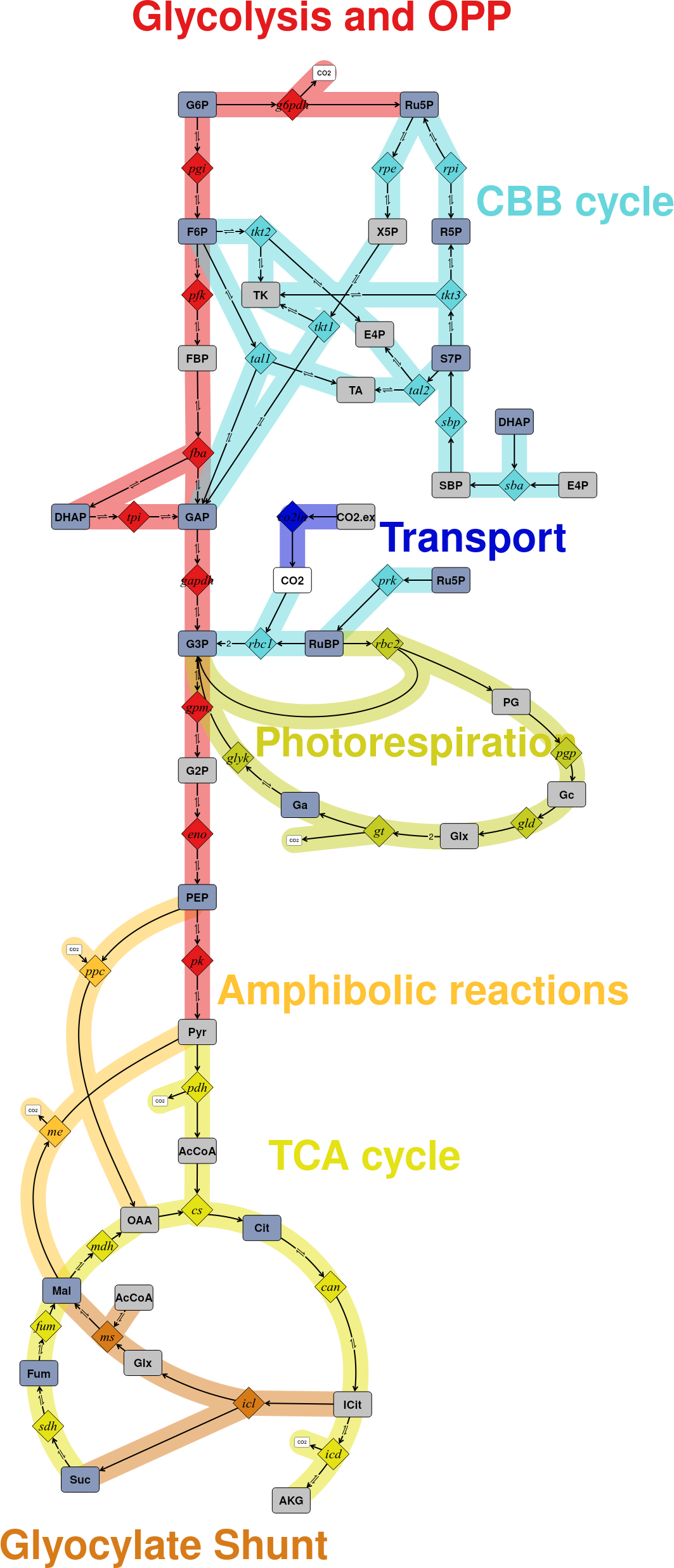}
    \caption{\textbf{Network model of \Synfull{}}. Throughout the text it is referred to as \Syn.}
    \label{fig:syn}
\end{figure}

\noindent \Syn{} is a model of the central carbon metabolism of \Synfull. The model is provided with the \texttt{freeflux} software (\texttt{models/synechocystis}), being a variant for INST \cmfa{}. It comprises 57 reactions connecting 38 metabolites, features 992 isotopomers, and has 7/22/32 free model parameters (net fluxes/exchange fluxes/pool sizes). 
Two measurement configurations are available:
the original measurement configuration, termed \Syn{\_a}, has 18 MS measurements of intermediates, taken at eight increasingly spaced measurement time points, resulting in a total of 590 independent labeling measurements.
A second measurement configuration, termed \Syn{\_b}, has one MS measurement of glyceraldehyde-3-phosphate (G3P) at the same measurement time points as \Syn{\_a}, resulting in a total of 24 independent labeling measurements.
\clearpage
\bibliography{reference}
\end{document}